\documentclass[]{JHEP3}

\usepackage{amsmath,amssymb}
\usepackage{graphicx}

\newcommand \bra[1]{\left< {#1} \,\right\vert}
\newcommand \ket[1]{\left\vert\, {#1} \, \right>}

\newcommand{\bea}{\begin{eqnarray}}
\newcommand{\eea}{\end{eqnarray}}
\newcommand{\simgt}{\hbox{ \raise3pt\hbox to 0pt{$>$}\raise-3pt\hbox{$\sim$} }}
\newcommand{\simlt}{\hbox{ \raise3pt\hbox to 0pt{$<$}\raise-3pt\hbox{$\sim$} }}

\newcommand{\clfn}{\setcounter{footnote}{0}}

\newcommand{\nn}{\nonumber}

\newcommand{\dsl}[1]{\hspace{#1}/}
\newcommand{\mt}{m_{t}}
\newcommand{\mb}{m_{b}}
\newcommand{\mW}{m_{W}}
\newcommand{\Gt}{\Gamma_{t}}
\newcommand{\GW}{\Gamma_{W}}

\newcommand{\qqb}{q\bar{q}}
\newcommand{\ttb}{t\bar{t}}
\newcommand{\bwbw}{bW^{+}\bar{b}W^{-}}
\newcommand{\mtt}{m_{\ttb}}

\newcommand{\cM}{{\cal M}}
\newcommand{\cP}{{\cal P}}
\newcommand{\cD}{{\cal D}}
\newcommand{\cS}{{\cal S}_{F}}
\newcommand{\cSb}{\bar{\cal S}_{F}}

\newcommand{\cG}{{\cal G}}
\newcommand{\bt}{\beta_{t}}

\newcommand{\tsum}{{\textstyle \sum}}
\newcommand{\tb}{\bar{t}}
\newcommand{\bb}{\bar{b}}
\newcommand{\dilep}{\ell\ell}

\title{
Bound-state effects on kinematical distributions \\
of top quarks at hadron colliders}

\author{
Yukinari Sumino \\
Dept.\ of Physics, Tohoku University, Sendai 980-8578, Japan \\
E-mail: \email{sumino@tuhep.phys.tohoku.ac.jp}}

\author{
Hiroshi Yokoya \\
Theory Unit, Physics Department, CERN, CH-1211 Geneva, Switzerland \\
E-mail: \email{hiroshi.yokoya@cern.ch}}

\abstract{
First we present a theoretical framework to compute the
fully differential cross sections for the top-quark productions and
their subsequent decays at hadron colliders, incorporating the
bound-state effects which are important in the $\ttb$ threshold region.
We include the bound-state effects such that the cross sections are
correct in the LO approximation both in the threshold and high-energy
regions.
Then, based on this framework we compute various kinematical
distributions of top quarks as well as of their decay products at the
LHC, by means of Monte-Carlo event-generation.
These are compared with the corresponding predictions based on
conventional perturbative QCD.
In particular, we find a characteristic bound-state effect on the
$(bW^+)$-$(\bar{b}W^-)$ double-invariant-mass distribution, which is
deformed to the lower invariant-mass side in a correlated manner.
}

\keywords{Hadronic Colliders, Heavy Quark Physics, QCD}
\preprint{TU-867 \\ CERN-PH-TH-2010-147 \\ June 2010}

\begin{document}

\section{Introduction}

At the CERN Large Hadron Collider (LHC), the top quark will be produced
copiously.
The cross section for the top-quark pair-production amounts to several
hundred pb~\cite{Moch:2008qy,Cacciari:2008zb,Kidonakis:2008mu}, and
order $10^6$ top-quark events will be observed each year if the LHC runs
with 14~TeV collision energy and achieves the designed luminosity.
Collecting these top-quark events, detailed analysis on the properties
of the top quark will be possible, such as precise determinations of its
mass and width, structure of electroweak and strong interactions, and
its spin properties~\cite{Bernreuther:2008ju}.
The current world average of the top-quark mass measurements from
the combined analysis of CDF and D0 collaborations at the Fermilab
Tevatron reads $\mt=173.1\pm1.3$ GeV~\cite{:2009ec} (see also
\cite{Amsler:2008zzb}).
Furthermore, the top-quark production process is considered as a {\it
standard candle} process.
Namely, it serves understanding detector
performances, e.g.\ jet energy calibrations, from comparisons of
experimental measurements with
theoretically reliable or well-controllable predictions,
for observables including jet
topologies, backgrounds and underlying events.

There have been many studies on top-quark production
processes at the
LHC.\footnote{%
See e.g.\ \cite{Bernreuther:2008ju,Bernreuther:2010ny} for more complete
review.}
Update analyses on the total pair-production cross-section are presented
in \cite{Moch:2008qy,Cacciari:2008zb,Kidonakis:2008mu}, including
the next-to-leading-order (NLO)
correction~\cite{Nason:1987xz,Beenakker:1988bq} and resummation
of threshold
logarithms~\cite{Catani:1996dj,Kidonakis:1997gm,Bonciani:1998vc} in
QCD.
Differential distributions including decays of 
the top-quark have also been
known up to
NLO~\cite{Mangano:1991jk,Frixione:1995fj,Bernreuther:2004jv}, and
various distributions are investigated in
\cite{Frederix:2007gi,Melnikov:2009dn,Mahlon:2010gw,Bernreuther:2010ny,%
Biswas:2010sa}.

Recently, $\ttb$ invariant-mass ($\mtt$) distribution near threshold
has been investigated incorporating the bound-state
effects~\cite{Hagiwara:2008df,Kiyo:2008bv}.
The effects are found to be significant at the LHC, since (in contrast
to the Tevatron) the gluon-fusion channel dominates the cross section
and there are significant contributions from (the remnant of) the
color-singlet $\ttb$ resonances.

In this paper we compute the fully differential cross sections for the
top-quark pair productions and their subsequent decays at the LHC.
In particular, we incorporate the bound-state effects, which are
important in the $\ttb$ threshold region, into the cross sections.
We extend the studies of \cite{Hagiwara:2008df,Kiyo:2008bv} and present
a theoretical frame to incorporate the bound-state effects to the
differential cross sections.
Using this result, we compute various kinematical distributions of the
top quarks and their decay products at hadron colliders, by developing a
Monte-Carlo (MC) event-generator incorporating the bound-state effects.
(There exist similar MC event-generators for computing the top-quark
cross-sections in the $\ttb$ threshold region at future $e^+e^-$
colliders~\cite{Fujii:1993mk,Martinez:2002st,Ikematsu:2003pg}.)
Through the analysis, we elucidate the nature of the bound-state effects
at various stages: at the partonic matrix-element level, both with and
without including the decay of the top quark, and in the kinematical
distributions after incorporating the initial-state radiation (ISR)
effects.
Theoretically the fully differential cross sections contain more
information on the bound-state effects than just the $\ttb$
invariant-mass distribution; for instance, it is known that the top
momentum distribution is sensitive to the resonance wave functions in
momentum space~\cite{Sumino:1992ai,Jezabek:1992np}.
From a practical point of view, the differential cross sections are
useful for studying effects of various kinematical cuts, detector
acceptance corrections, detector calibrations, etc.

The method for incorporating $\ttb$ bound-state effects has been
developed mainly in the studies of $\ttb$ productions in $e^+e^-$
collisions~\cite{Fadin:1987wz,Fadin:1988fn,Hoang:2000yr}.
Formally, in the limit where we neglect the top-quark width,
$\Gt\to0$, bound-state effects can be incorporated by resummation
of the Coulomb singularities $(\alpha_s/\beta)^n$, where $\beta$ is the
velocity of the top quark in the $\ttb$ c.m.\ frame.
In contrast to the $e^+e^-$ collision, at hadron colliders, $\ttb$ pairs
are produced in both color-singlet and octet states, and the (partonic)
collision energy is not fixed.
Due to the latter reason, we have to set up a theoretical framework
which is valid both in the threshold region ($\mtt\simeq2\mt$) and in
the high-energy region ($\mtt\gg 2\mt$).
The former region is where the bound-state effects (Coulomb corrections)
become significant and where the non-relativistic approximation is valid.
On the other hand, in the latter region, the bound-state effects are
not significant and the top quarks are relativistic.
We present a framework which takes into account all the leading-order
(LO) corrections in both regions.
Namely, we incorporate all the $(\alpha_s/\beta)^n$ terms in the
threshold region, while we include all the $\beta^n$ terms in the
relativistic region.
(Some of the important subleading corrections are also incorporated.)
Furthermore, we interpolate the two regions smoothly in a natural way.

Another important aspect in computing the differential cross sections
for the top-quark productions and decays is to construct full
amplitudes corresponding to the $\bwbw$ final state and to incorporate
off-shellness of the top quarks appropriately~\cite{Kauer:2002sn}.
The former is important to incorporate the polarizations of $t$ and
$\bar{t}$ and angular correlations in their decay products.
(In Appendix D, we will further discuss the polarizations of the $W$'s
from the top quarks.)
The latter is intimately related to the former and is known to be
particularly important in the threshold region.
In the $\bwbw$ production, there are non-resonant diagrams where $bW^+$
and $\bb W^-$ are not produced from the decay of $t$ and $\tb$.
Often the non-resonant diagrams are omitted in the studies of $\ttb$
productions, since they are suppressed in the events where both $t$ and
$\tb$ are nearly on-shell.
In the threshold region, however, either of $t$ and $\tb$ tends to be
off-shell due to restricted phase-space and the
binding effects~\cite{Sumino:1992ai}, and the
non-resonant diagrams can give non-negligible contributions compared to
the resonant ($\ttb$) diagrams~\cite{Beneke:2010mp}.
Since these contributions interfere with each other, all the diagrams
have to be taken into account at the amplitude level.
Moreover, because of the requirement by unitarity, we also have to
ensure a consistent  treatment of the finite decay width of the top
quark in our framework.
We will discuss these points within our framework, which includes the
bound-state effects as well as the non-resonant diagrams, in connection
with a known problem regarding a gauge cancellation.

In order to compute numerically various kinematical distributions at
hadron colliders, we develop a MC event-generator,
which is adapted to the {\tt MadEvent}~\cite{Maltoni:2002qb,Alwall:2007st}
environment\footnote{
The Fortran code for the event generator including the bound-state
corrections is available at
{\tt http://madgraph.kek.jp/\~{}yokoya/TopBS/}.}. 
We include the bound-state effects in the hard-scattering part of the LO
event-generator, on the basis of our theoretical framework.
The ISR and/or final-state radiation (FSR), which are of importance at
hadron colliders, are incorporated via the parton-shower approach.
Since the parton shower does not alter the normalizations of the cross
sections at the partonic level, we will complement the overall
normalizations, known up to NLO~\cite{Hagiwara:2008df,Kiyo:2008bv}, by
multiplying the cross sections in the individual channels with the
so-called ``$K$-factors.''
We note, however, that 
our aim here is to construct a generator valid only up to
LO with respect to the differential distributions, in this first attempt
to include the bound-state effects.
Compare with the existing NLO event-generators, such as {\tt
MC@NLO}~\cite{Frixione:2002ik,Frixione:2003ei} and {\tt
POWHEG}~\cite{Frixione:2007nw}, which realize a consistent treatment of
perturbative corrections for any process and any phase-space point.

Using the generated events, we study the bound-state effects on the top
quark differential distributions at the LHC, focusing on the events in a
relatively low $\mtt$ region.
A $\ttb$ pair gains a binding energy due to exchange of Coulomb gluons
between them.
This effect tends either of $t$ and $\tb$ to be off-shell below the
threshold, and the effect remains even a few tens
GeV above the threshold, due
to the large width of the top quark.
We will quantify this picture through detailed examinations of the top
quark differential distributions.
\medbreak

The paper is organized as follows.
In Sec.~2, we give a theoretical framework for computing the amplitudes
for top-quark pair-production at hadron colliders, incorporating the
bound-state effects (Sec.~2.1), the finite width effects (Sec.~2.2), and
the ISR effects and $K$-factors (Sec.~2.3).
In Sec.~3, we present numerical studies for various kinematical
distributions in $\ttb$ production, using the MC simulation which
implements the ingredients explained in the previous section.
In Sec.~4, we summarize our results.
To avoid complexity in the main body of the paper, several detailed
discussions are presented in the Appendices.
In App.~A, we identify the $\ttb$ Green function in a Feynman amplitude.
In App.~B, we derive the off-shell suppression factor.
In App.~C, the color decomposition of the amplitude is explained.
In App.~D, we examine the leptonic decays of $W$'s from top quarks with
and without spin correlations.

\section{Inclusion of Bound-state Effects}

In this section we present a theoretical
investigation of how to include the $\ttb$ bound-state
effects in the matrix elements for $gg\to\bwbw$
and $\qqb\to\bwbw$.
In particular we include the effects such that
the amplitude is correct in the leading-order approximation
both in the $\ttb$ threshold region and in the high energy
region.
Inclusion of several different effects is explained in steps:
In Sec.~2.1 we explain how to
incorporate the bound-state effects;
in Sec.~2.2 important higher-order effects of the large top-quark
decay-width are incorporated;
in these subsections, we consider only the partonic $S$-matrix
elements.
In Sec.~2.3 we incorporate the ISR effects and the $K$-factors
in the
corresponding partonic differential cross sections.

For later convenience, we divide each
amplitude into two
parts, the $\ttb$ (double-resonant) part and 
the non-resonant part, as
\begin{align}
 & \cM^{(c)}(I\to\bwbw) = \cM^{(c)}_{\ttb}(I\to\ttb\to\bwbw)
 + \cM^{(c)}_{\rm nr}(I\to\bwbw)\,,
 \label{eq:mesum}
\end{align}
where
$I=gg$ or $\qqb$ represents initial-state partons, and 
$c$ represents the color ($c=1$ and 8 for the singlet and
octet, respectively) of $I$, or equivalently, of 
$b\bb$ in the final-state.
The first term on the right-hand side
represents the sum of the diagrams which
contain both $t$ and $\tb$ as an intermediate state.
This part of the amplitude consists of $I\to\ttb$ processes followed
by subsequent decays of $t$ and $\tb$.
The second term represents the sum of the rest of the diagrams,
which consists of 
single(-top)-resonant diagrams and non-resonant diagrams.
Fig.~\ref{fig:ggqqtt} shows the tree-level
Feynman diagrams for 
the processes $gg\to\ttb$ and
$\qqb\to\ttb$.
\FIGURE[t]{
\includegraphics[width=.8\textwidth,clip]{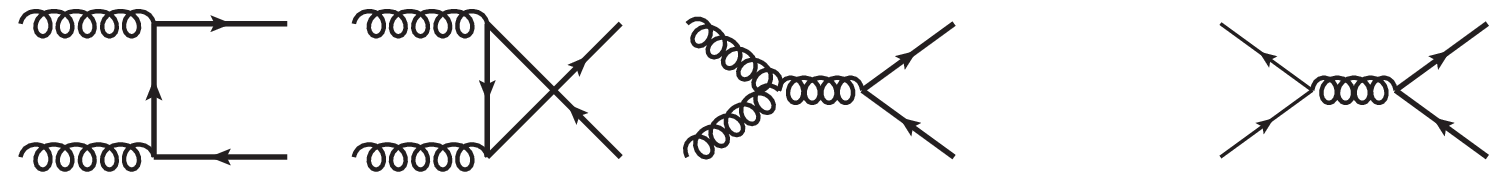}
\hspace{30pt}
\caption{Feynman diagrams for $gg\to\ttb$ and $\qqb\to\ttb$ at the
 tree-level. \label{fig:ggqqtt}}
}
Some examples of the tree-level diagrams included in
each part for $I=gg$ are shown in Fig.~\ref{fig:dia}.
\FIGURE[t]{
\includegraphics[width=.17\textwidth,clip]{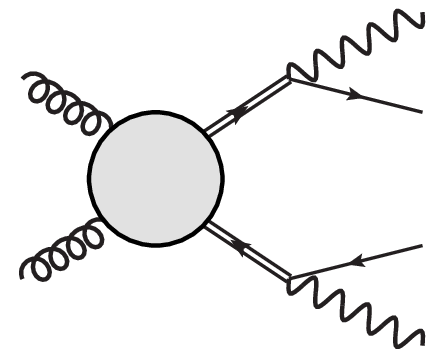}
\hspace{30pt}
\includegraphics[width=.17\textwidth,clip]{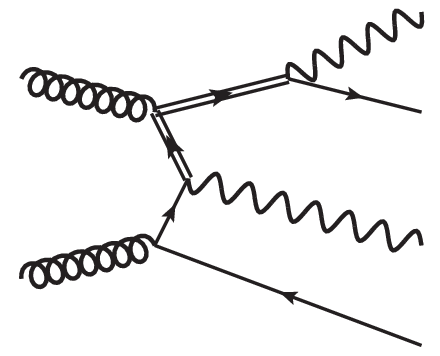}
\hspace{30pt}
\includegraphics[width=.17\textwidth,clip]{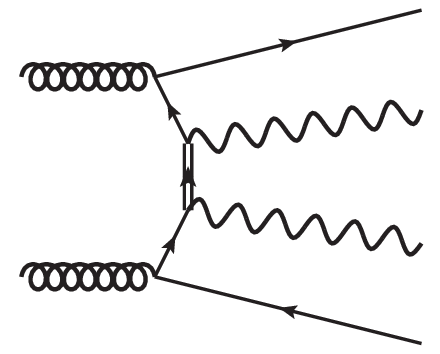}
\caption{Typical Feynman diagrams for the double-resonant (left),
 single-resonant (middle) and non-resonant (right) contributions 
 in the $gg\to\bwbw$ process.
  They belong to $\cM^{(c)}_{\ttb}$, $\cM_{\rm nr}^{(c)}$ and
  $\cM_{\rm nr}^{(c)}$, respectively.
 The term
 ``resonant'' is used to refer to the $t$ or $\tb$ quark
 propagator (shown with double line) which can become close to on-shell.
 The blob in the left diagram represents the first three diagrams in
 Fig.~\ref{fig:ggqqtt}.\label{fig:dia}}
}
In general each part is gauge-dependent.
In this paper, we work in Feynman gauge for $SU(3)_{\rm c}$ and in
unitary gauge for the broken electroweak symmetry. 

In computing the
tree-level Feynman diagrams which contain the top-quark
propagators,
we include the (on-shell) top-quark decay-width $\Gt$ in the
propagator denominator  as
\bea
&&
 \cS(p_{t})=
 \frac{i(p\dsl{-5pt}_{t}+\mt)}{p_{t}^{2}-\mt^{2}+i\mt\Gt}
\,.
\label{eqFWS}
\eea

\subsection{LO cross section valid from threshold to high energies}

In this subsection we include the bound-state effects in the $\ttb$
amplitude $\cM_{\ttb}^{(c)}$.
We consider the narrow-width limit of top-quark width in this
subsection.
Namely, we take into account only the leading contributions as
$\Gt\to 0$.
Important subleading effects by the finite top-quark width will be
investigated separately in the next subsection.
\medbreak

We start by reviewing the conventional method for including the
bound-state effects in the matrix element (or in the fully differential
cross section) of $e^+e^-\to\ttb\to\bwbw$ close to the
threshold of $\ttb$ pair productions.
At the leading-order, 
this is achieved by multiplying the tree-level
amplitude corresponding to the diagrams
$e^+e^-\to\ttb\to\bwbw$ by an enhancement factor as
\cite{Sumino:1992ai,Jezabek:1992np}\footnote{%
There are two tree-level diagrams for $e^+e^-\to\ttb\to
\bwbw$ with $\gamma$ and $Z$ boson intermediate states.
$\cM(e^+e^-\to\ttb\to\bwbw)_{\rm tree}$ denotes the sum of
them.
}
\bea
\cM(e^+e^-\to\ttb\to\bwbw) =
\cM(e^+e^-\to\ttb\to\bwbw)_{\rm tree} 
\nonumber\\
\times
\frac{G^{(1)}(E+i\Gt,\vec{p})}{G_{0}(E+i\Gt,\vec{p})}.
\label{NR-enhancement}
\eea
Here, the non-relativistic Green functions are defined by\footnote{
In this study, 
$G^{(c)}(E+i\Gt,\vec{p})$ is computed 
numerically by solving the Schr\"odinger
equation in coordinate space and taking Fourier transform 
\cite{Sumino:1992ai}.
Alternatively, one may
solve the Schr\"odinger equation in momentum space directly
\cite{Jezabek:1992np}.
}
\bea
&&
\left[ (E+i\Gt)-
\left\{ - \frac{\nabla^2}{\mt} + V_{\rm QCD}^{(c)}(r)\right\}\right]
\tilde{G}^{(c)}(E+i\Gt,\vec{r}) = \delta^3(\vec{r}) ,
\label{defGreenfn}
\\&&
G^{(c)}(E+i\Gt,\vec{p})=\int d^3\vec{r}\, e^{-i\vec{p}\cdot \vec{r}}\,
\tilde{G}^{(c)}(E+i\Gt,\vec{r})\,.
\eea
$E=\sqrt{s}-2\mt$ is the c.m.\ energy measured from the threshold;
$\mt$ is the pole mass of the top quark;
$\vec{r}$ denotes the relative coordinate of $t$ and $\tb$,
while $\vec{p}$ denotes the three-momentum of $t$ (or minus the
three-momentum of $\tb$), both defined in the c.m.\ frame;
$V_{\rm QCD}^{(c)}(r)$ is the QCD potential between the
$\ttb$ pair in the color-singlet ($c=1$) or color-octet
($c=8$) channel.
In $e^+e^-$ collisions, $\ttb$ pairs are produced in the color-singlet 
channel, hence $c=1$ in Eq.~(\ref{NR-enhancement}).
The free non-relativistic Green function
$G_{0}(E+i\Gt,\vec{p})$ is obtained from $G^{(c)}(E+i\Gt,\vec{p})$
by setting $V_{\rm QCD}^{(c)}(r)\to 0$.
Formally the above Green function can be expressed as
\bea
G^{(c)}(E+i\Gt,\vec{p}) =
\bra{\vec{p}\rule[-2mm]{0mm}{6mm}}
\frac{1}{E+i\Gt-\vec{p}^{\,2}/\mt-V_{\rm QCD}^{(c)}(r)}
\ket{\vec{r}=\vec{0}} ,
\label{NRGreenFn}
\eea
using an operator notation in quantum mechanics.
By definition, the above Green function contains only the
$S$-wave contributions.

There are two methods to compute the
total cross section for $e^+e^-\to\ttb\to\bwbw$
incorporating the bound-state effects.
One way is to integrate the absolute square of the matrix element given
in Eq.~(\ref{NR-enhancement}) over the phase-space of the final
$\bwbw$ state.
The other method is to use the optical theorem (unitarity relation) and
take the imaginary part of a current-current correlator (two-point
function).
At the leading-order, the latter method 
leads to the formula~\cite{Fadin:1987wz,Fadin:1988fn}:
\bea
\sigma_{\rm tot}(e^+e^-\to\ttb\to\bwbw)
=
\sigma_{\rm tot}(e^+e^-\to\ttb)_{\rm tree}
~~~~~~~~~~~~~~~~
\nonumber\\
\times
\frac{\mathop{\rm Im}\Bigl[ \tilde{G}^{(1)}(E+i\Gt,\vec{r}=\vec{0}) \Bigr]}
{\mathop{\rm Im} \Bigl[ \tilde{G}_{0}(E,\vec{r}=\vec{0}) \Bigr]},
\label{NR-totcs-LO}
\eea
where the Green function in coordinate space is defined 
in Eq.~(\ref{defGreenfn}).
$\sigma_{\rm tot}(e^+e^-\to\ttb)_{\rm tree}$ denotes the Born
cross-section for the production of on-shell top quarks.
Note that in the denominator we set $\Gt$ to zero
in $\tilde{G}_{0}$, whereas $\Gt$ is retained in
the denominator in Eq.~(\ref{NR-enhancement}).
The different treatment of $\Gt$ is because in Eq.~(\ref{NR-enhancement})
we use 
the tree-level amplitude with {\it unstable top quarks} 
(in the intermediate state), whereas
in Eq.~(\ref{NR-totcs-LO}) we use the
tree-level cross section of the {\it on-shell top quarks}
(in the final state).

Both formulas (\ref{NR-enhancement}) and (\ref{NR-totcs-LO}) incorporate
all the leading-order corrections $\sim (\alpha_s/\beta)^n$
in the threshold region $E\ll \mt$.
On the other hand, the formulas are not valid at higher c.m.\
energies
$E\simgt \mt$, since relativistic corrections $\sim \beta^n$,
which grow with energy, are neglected in these formulas.
\medbreak

Now we turn to the partonic cross sections for the
top-quark productions in hadron collisions,
$I\to\ttb\to\bwbw$ with $I=gg$ and
$\qqb$.
Unlike $e^+e^-$ collisions, the collision energy of
the initial state cannot be fixed.
Thus, we need to consider both threshold and high-energy regions.
The formulas which we propose, valid in both regions
within the leading-order approximation, can be summarized as follows:
\\
\begin{itemize}
\item[(1)]
The $\ttb$ amplitude for $I\to\ttb\to\bwbw$
is given by
\bea
\cM^{(c)}_{\ttb}(I\to\ttb\to\bwbw) &=&
\cM^{(c)}_{\ttb}(I\to\ttb\to\bwbw)_{\rm tree}
\nn\\&&~~~~~~~~~~~~~~~~~
\times
\frac{G^{(c)}(E'+i\Gt,\vec{p})}{G_0(E'+i\Gt,\vec{p})} ,
\label{enhancement2}
\eea
with
\bea
E'=E+\frac{E^2}{4\mt} \,.
\label{eq:ene}
\eea
The Feynman diagrams which contribute to the tree-level amplitude
$\cM^{(c)}_{\ttb}(I\to\ttb\to\bwbw)_{\rm tree}$ are those shown 
in Fig.~\ref{fig:ggqqtt}, after attaching the decay
vertices $t\to bW^+$ and $\bar{t}\to\bar{b}W^-$ to each
diagram.
There are
both color-singlet ($c=1$) and color-octet ($c=8$)
channels in the case $I=gg$, while there is only the
color-octet channel in the case $I=\qqb$.
Here, $E$ is defined from the
$\ttb$ invariant-mass $\mtt$ as
$E=\mtt-2\mt$.
The only essential difference from the corresponding
formula for the $e^+e^-$ collision,
Eq.~(\ref{NR-enhancement}),
is the use of the modified energy
$E'$ [Eq.~(\ref{eq:ene})] instead of $E$.
\\
\item[(2)]
We may compute the $\ttb$ invariant-mass distribution by
integrating the absolute square of the above amplitude
$|\cM^{(c)}_{\ttb}(I\to\ttb\to\bwbw)|^2$ over the
$\bwbw$ phase-space for each fixed $\mtt$.
Alternatively we may obtain the formula for the $\ttb$ invariant-mass
 distribution using the optical theorem, similarly to the $e^+e^-\to
\ttb$ case.
In the leading-order approximation, this reads
\bea
&&
\hat{\sigma}_{\rm tot}^{(c)}(I\to\ttb\to\bwbw)
=
\hat{\sigma}_{\rm tot}^{(c)}(I\to\ttb)_{\rm tree}
\times
\frac{\mathop{\rm Im}\Bigl[ \tilde{G}^{(c)}(E'+i\Gt,\vec{r}=\vec{0}) \Bigr]}
{\mathop{\rm Im} \Bigl[ \tilde{G}_{0}(E',\vec{r}=\vec{0}) \Bigr]} \,.
\nn\\
\label{totcs-LO}
\eea
As in the $e^+e^-\to\ttb$ case, $\hat{\sigma}_{\rm
tot}^{(c)}(I\to\ttb)_{\rm tree}$ is the Born cross-section for the on-shell
top quarks; accordingly $\Gt$ is set to zero in
$\tilde{G}_0(E',\vec{r}=\vec{0}\bigr)$.
We note that the $\ttb$ invariant-mass distribution obtained from
Eq.~(\ref{totcs-LO}) does not exactly coincide with that obtained by
integrating $|\cM^{(c)}_{\ttb}(I\to\ttb\to\bwbw)|^2$
over the $\bwbw$ phase-space; the difference is ${\cal
O}(\Gt/\mt)$
and will be discussed in the next subsection. 
\end{itemize}

In the following we sketch our theoretical consideration which led to the
above formulas (\ref{enhancement2}) and (\ref{totcs-LO}), where
some details are relegated to App.~\ref{app:green}.
As explained in that appendix, part of the Feynman amplitude
for $I\to\ttb\to\bwbw$ can be identified with a Green function that
dictates the time evolution of the $\ttb$ system.
(Such an identification is possible in all kinematical regions.)
In the c.m.\ frame of $\ttb$, for the initial-state $\ket{i}$
and final-state $\ket{f}$ of the $\ttb$ system,
this Green function can be written formally as\footnote{
As already stated, treatment of the top-quark width
is correct only in the leading order.
}
\bea
\bra{f}
\frac{1}{\mtt-H+i\Gt} \ket{i},
\label{RelGreenFn}
\eea
where
$\mtt$ is the c.m.\ energy of $\ttb$
($\ttb$ invariant-mass).
The full QCD Hamiltonian is denoted by $H$.
Because of this property, the amplitude for $I\to\ttb\to
\bwbw$, which incorporates the $\ttb$ bound-state effects,
can be obtained from the tree-level amplitude
by multiplying an enhancement factor:
\bea
\cM^{(c)}_{\ttb}(I\to\ttb\to\bwbw) \approx
\cM^{(c)}_{\ttb}(I\to\ttb\to\bwbw)_{\rm tree} 
~~~~~~~~
&&
\nn\\
\times
\frac{
\displaystyle
\bra{\vec{p}\rule[-2mm]{0mm}{6mm}}\frac{1}{\mtt-H+i\Gt}
\ket{\vec{r}=\vec{0}}
}{\displaystyle
\bra{\vec{p}\rule[-2mm]{0mm}{6mm}}\frac{1}{\mtt-H_0+i\Gt}
\ket{\vec{r}=\vec{0}}
} 
.
&&
\label{enhancement}
\eea
$H_0$ denotes the Hamiltonian $H$ after setting $\alpha_s\to 0$,
i.e.\ the free Hamiltonian.
As before,
$\vec{r}$ denotes the relative coordinate of $t$ and $\tb$,
while $\vec{p}$ denotes the three-momentum of $t$, both defined
in the c.m.\ frame of the $\ttb$  system.
Corrections to Eq.~(\ref{enhancement}), which 
come from the non-resonant part of the amplitude (i.e.\ that
vanish as $\Gt \to 0$), are neglected;
see Appendix~\ref{app:green}.

In the above equation, we have taken advantage of the fact that we work
in the leading-order approximation and set the
initial-state of the Green functions as $|\vec{r}=\vec{0}\rangle$.
This follows from the following consideration.
Naively, the
$\ttb$ system cannot be regarded as being created 
by contact interaction
(i.e. at the same point $\vec{r}=\vec{0}$) in 
the $t$- and $u$-channel diagrams of $gg\to\ttb$, in which 
the top-quark
is exchanged between the initial $gg$ (see Fig.~\ref{fig:ggqqtt}).
In the non-relativistic
region, however, we may set $\vec{p}\to 0$ in
the $t$- or $u$-channel top-quark propagator in the leading-order
approximation.
The denominator of the propagator effectively reduces to a constant
close to the threshold, so that 
$\ttb$ can be regarded as being created at the same point.
On the other hand, in the relativistic region the enhancement factor in
Eq.~(\ref{enhancement}) reduces to $1+{\cal O}(\alpha_s)$.
Hence, it is justified to evaluate the Green functions with the
initial-state $|\vec{r}=\vec{0}\rangle$ within our present
approximation.

Since our aim is to include the leading-order contributions both
in the relativistic and non-relativistic regions, the full form
of the Hamiltonian is not necessary.
In the region where $t$ and $\tb$ are relativistic,
$\mtt-2\mt\simgt \mt$, the leading-order contribution in the
Hamiltonian reads
\bea
H = 2 \sqrt{\vec{p}^{\,2}+\mt^2} + {\cal O}(\alpha_s) \,,
\label{relH}
\eea
as shown
in Appendix~\ref{app:green}.
It
is nothing but the sum of the energies of
free on-shell $t$ and $\tb$.
The above equation also indicates how the next-to-leading order effects
enter the Hamiltonian.
On the other hand, in the non-relativistic region,
$E=\mtt-2\mt\ll \mt$, the leading-order contributions in the
Hamiltonian can be written explicitly as
\bea
H = 2\mt + \Biggl[\frac{\vec{p}^{\,2}}{\mt} + V_{\rm QCD}^{(c)}(r)
\Biggr]\times\Bigl[ 1
+{\cal O}(\alpha_s,\beta)\Bigr].
\eea
It is indicated that the next-to-leading order corrections
enter as ${\cal O}(\beta)$ relativistic corrections or ${\cal O}(\alpha_s)$
corrections.

A natural choice of the Hamiltonian, which incorporates the
leading-order contributions in both regions and smoothly
interpolates these regions, is given by
\bea
H_{\rm LO} = 2 \sqrt{\vec{p}^{\,2}+\mt^2}  + V^{(c)}_{\rm QCD}(r).
\label{H-naturalchoice}
\eea
In fact, it is well known that,
when computing higher-order corrections in Coulombic bound-state
problems, part of them (relativistic corrections)
contribute exactly in the above
form~\cite{LL}.
Thus, in principle,
one may determine the enhancement factor
in Eq.~(\ref{enhancement}) using the above Hamiltonian.

Due to technical reasons, however,
we use an alternative
form of the enhancement factor, which is equivalent
within the present approximation. 
By substituting
$E=\mtt-2\mt$
to the on-shell relation
\bea
\mtt= 2 \sqrt{\vec{p}^{\,2}+\mt^2} ,
\label{onshell-cond}
\eea
one finds that
\bea
E+\frac{E^2}{4\mt}=\frac{\vec{p}^{\,2}}{\mt}.
\label{onshell-cond2}
\eea
Therefore, if we define
\bea
&&
G^{(c)}\bigl(E'+i\Gt,\vec{p}\bigr)=
\bra{\vec{p}\rule[-2mm]{0mm}{6mm}}\frac{1}
{E'+i\Gt-(\vec{p}^{\,2}/\mt+V^{(c)}_{\rm QCD}(r))}
\ket{\vec{r}=\vec{0}}
\label{mod-Greenfn}
\eea
with $E'$ defined by Eq.~(\ref{eq:ene}),
the position of the pole of $G^{(c)}(E'+i\Gt,\vec{p})$ is the same as
that of 
$\langle \vec{p}|(\mtt-H_{\rm LO}
+i\Gt)^{-1}|\vec{r}=\vec{0}\rangle$ 
in the limit $\alpha_s\to 0$.
Furthermore,  in the non-relativistic region,
evidently $G^{(c)}(E',\vec{p})$ is the same as
$\langle \vec{p}|(\mtt-H_{\rm LO}
+i\Gt)^{-1}|\vec{r}=\vec{0}\rangle$
in the leading-order approximation.
Hence, we may compute the matrix element by
the formula Eq.~(\ref{enhancement2}).
One may be worried that, although the replacement $E\to E'$
correctly accounts for the pole position, 
the residue of the pole may be
altered significantly from that of Eq.~(\ref{enhancement})
in the relativistic region.
This is not the case, since the change of the residue
will be canceled in the ratios of the Green functions
in  Eqs.~(\ref{enhancement2}) and (\ref{totcs-LO}).

The advantage of using $G^{(c)}(E'+i\Gt,\vec{p})$ is that one can
obtain it from the conventional non-relativistic Green function
with a minimal modification $E \to E'=E+{E^2}/({4\mt})$.
In particular, properties of $G^{(c)}$ are 
fairly well known.

Let us comment on 
the dependence of the 
Green function on the top-quark width $\Gt$ 
in Eq.~(\ref{mod-Greenfn}).
In Eq.~(\ref{eqFWS})
the shift of the pole position of the top-quark
propagator due to the finite top-quark width
can be incorporated simply by a replacement
$\mt^2\to \mt^2-i\mt\Gt$.
If we apply it to Eq.~(\ref{onshell-cond}),
one finds that $i\Gt$ will be added to the left-hand side
of Eq.~(\ref{onshell-cond2}).
Hence, inclusion of $\Gt$ as in
Eq.~(\ref{mod-Greenfn}) is correct in the
leading-order approximation.

Throughout our analysis, we include an important subleading correction
to the bound-state effects, in order to
make our analysis more realistic.
This is the NLO (1-loop) correction to the static potentials
$V^{(c)}_{\rm QCD}(r)$ between the $\ttb$ pair.
The NLO potential reads~\cite{Kniehl:2004rk}
\bea
V_{\rm QCD}^{(c)}(r;\mu_{\rm B})=
C^{(c)} \frac{\alpha_s(\mu_{\rm B})}{r}
\left[
1 + \frac{\alpha_s(\mu_{\rm B})}{4\pi}
\left\{ 2\beta_0\left[\ln(\mu_{\rm B} r) + \gamma_E\right]+ a_1^{(c)}
\right\}
\right]
\eea
with
\bea
&&
C^{(1)}=-C_F,
~~~~
C^{(8)}=\frac{C_A}{2}-C_F ,
\\&&
\beta_0=\frac{11}{3}C_A-\frac{2}{3}n_q,
~~~~~
a_1^{(1)}=a_1^{(8)}= \frac{31}{9}C_A - \frac{10}{9}n_q,
\eea
for the $\overline{\rm MS}$ coupling.
Here, $\gamma_E=0.5772...$ denotes the Euler constant;
$C_F=4/3$ and $C_A=3$ are color factors.
The QCD potential is renormalization-group invariant,
and we evaluate the above expression at the Bohr scale of
$\mu_{\rm B}=20$~GeV and with $n_q=5$ ($\alpha_{s}(\mu_{B})=0.153$).
\\

Now we perform a few
tests of our formulas, Eqs.~(\ref{enhancement2}) and (\ref{totcs-LO}).
First we examine the impact of the replacement $E\to E'$.
The $\ttb$ invariant-mass distributions 
$d\hat{\sigma}/d\mtt$
computed with these formulas
are compared with $d\hat{\sigma}/d\mtt$ computed
by the formulas valid only in the threshold region,
namely Eqs.~(\ref{enhancement2}) and (\ref{totcs-LO}) after we replace
$E'$ by $E$.
\FIGURE[t]{
\includegraphics[width=.6\textwidth,clip]{sigh_ene_ggsng.eps}
\caption{Partonic $\ttb$ invariant-mass distributions  
for $gg\to\ttb\to\bwbw$ in the
 color-singlet channel.
The green solid line is calculated with $E'=E+E^2/(4\mt)$ 
[Eq.~(\ref{eq:ene})], while the red dot-dashed line is calculated 
with $E$ instead of $E'$.
The black dashed line represents the Born cross-section.
The dotted lines are those for the $gg\to\ttb$ process
(production of on-shell top quarks).
\label{fig:sigh_ene}}
}
In Fig.~\ref{fig:sigh_ene} we plot $\hat{\sigma}$ 
for $gg\to\ttb$ in the color-singlet channel.\footnote{
In Secs.~2.1 and 2.2, 
the partonic $\ttb$ invariant-mass distribution (before 
including the effects of ISR and
parton distribution function) 
is proportional to the partonic total
cross-section and delta function, 
$d\hat\sigma/d\mtt\propto\hat\sigma\delta{(\hat{s}-\mtt^2)}$;
c.f.\ eq.~(17) of \cite{Hagiwara:2008df}.
Hence, we plot $\hat{\sigma}(\hat{s}=\mtt^2)$ instead of $d\hat\sigma/d\mtt$
in these subsections.}
The green solid and dotted lines are those computed using
Eqs.~(\ref{enhancement2}) and (\ref{totcs-LO}), respectively, while the
red dot-dashed and dotted lines are those computed with the same
formulas but after the replacement $E'\to E$.
For comparison, the Born cross-sections [using Eqs.~(\ref{enhancement2})
and (\ref{totcs-LO}) but without the enhancement factors] are also
plotted with the dashed and dotted black lines.
All the cross sections are computed with $\mt=173$~GeV and
$\Gt=1.49$~GeV (the tree-level top-quark decay-width).

The difference between the $\ttb$ invariant-mass distributions using
Eqs.~(\ref{enhancement2}) and (\ref{totcs-LO}) (solid and dotted green
lines) is due to ${\cal O}(\Gt/\mt)$ corrections.
The replacement $E'\to E$ in Eq.~(\ref{totcs-LO}) changes the $\ttb$
invariant-mass distribution slightly above the $\ttb$ threshold;
compare the green dotted and red dotted lines.
The difference between the two cross sections is about 2.5\% in the
large $\mtt$ region.

On the other hand, the effect of the replacement $E'\to E$ in
Eq.~(\ref{enhancement2}) is much more pronounced above the $\ttb$
threshold.
There exist a large enhancement which amounts to nearly a factor of two
around $\mtt=400$~GeV; compare the green solid and red dot-dashed lines.
The origin of this large enhancement can be identified with a mismatch
of the on-shell conditions satisfied by the pole positions of the $t$
and $\tb$ propagators contained in $\cM^{(c)}_{\ttb,{\rm tree}}$ and by the
pole position contained in $G_0(E+i\Gt,\vec{p})$.
[Note that $\hat{\sigma}^{(c)}_{\rm tot,tree}$ in Eq.~(\ref{totcs-LO})
does not contain the $t$ or $\tb$ propagator, so that this
mismatch problem does not occur when we replace $E'$ by $E$ in
Eq.~(\ref{totcs-LO});
compare the green and red dotted lines.]
In fact,
the mechanism of this abnormally large deviation is closely tied to a
characteristic bound-state effect on the invariant-mass distributions of
the $bW^+$ and $\bb W^-$ systems.
We will investigate this issue in detail in Sec.~3, in which
we examine closely the differential distributions.
Nevertheless, even without going into these details,
the present comparison clearly shows the necessity of a proper treatment
of the relativistic kinematics, when we include the bound-state effects
to the fully differential cross section
of the process $I\to\ttb\to\bwbw$.

Since our formulas are correct in the narrow width limit
$\Gt\to 0$, the unitarity relation should be restored
in this limit.
In order to check this, in Fig.~\ref{fig:rt_360} we plot the ratios of 
$d\hat{\sigma}/d\mtt$ computed
using Eqs.~(\ref{enhancement2}) and (\ref{totcs-LO}) as we vary the
value of $\Gt$ at a fixed $\mtt$ of $360$~GeV.\footnote{
As we vary the value of $\Gt$ in the $t$ propagator and the Green
functions, the value of the weak gauge coupling constant $g_W$ in the
$tbW$ vertex is varied consistently, such that the tree-level top-quark
width takes the correct value.
}
\FIGURE[t]{
\includegraphics[width=.6\textwidth,clip]{rt_360.eps}
\caption{
Ratios of the partonic cross sections for
$gg\to\bwbw$ and  $gg\to\ttb$, 
in the color-singlet channel at
 ${{m}_{\ttb}}=360$~GeV.
These are plotted as functions of $\Gt$.
Green solid, red dashed and black dotted lines are those with 
the modified energy $E'$, with $E$ instead of $E'$,
 and for the Born cross-sections, respectively.
\label{fig:rt_360}}
}
We confirm that the ratio approaches to unity as $\Gt$ is reduced,
in the case that we use our relativistic formulas (green solid line) or
in the case that we use the tree-level cross sections (black dotted
line).
In sharp
contrast, the ratio does not approach to unity in the case that we
replace $E'$ by $E$ (red dashed line) due to the 
mismatch problem. 
It shows invalidity of the
non-relativistic approximation far above the threshold, especially for the
fully differential cross section. \\

As is well known, the leading-order bound-state effects in the
$\ttb$ threshold region are contained in the $S$-wave part of the
amplitude.
In the case of $gg\to\ttb$, the $S$-wave contributions reside in the
$J=0$ amplitude both for the color-singlet and color-octet channels.
Hence, it may be more appropriate to include the bound-state effects
only in the $J=0$ amplitude, rather than multiplying the whole 
$\ttb$ amplitude
by the enhancement factor as in Eq.~(\ref{enhancement2}).
Theoretically, the difference between the two prescriptions is
subleading.
We examine this feature by comparing the $\ttb$ invariant-mass 
distributions computed in both ways.
\FIGURE[t]{
\includegraphics[width=.6\textwidth,clip]{sigh_prs.eps}
\caption{Partonic $\ttb$ invariant-mass distributions 
for $gg\to\ttb$
in the color-singlet (red), color-octet
 (blue) channels, and the sum of them (black).
The solid lines are calculated by multiplying the bound-state
enhancement factor
 to the whole $\ttb$ amplitude.
The dashed lines are calculated as the
 sum of contributions from
all $J$'s, where only the $J=0$ amplitude is multiplied by the
 enhancement factor.
The dotted lines represent the Born cross-sections 
corresponding to the above lines.
\label{fig:prs}}
}
In Fig.~\ref{fig:prs}, we plot the $\ttb$ invariant-mass distributions
for $gg\to\ttb$ process.
Each solid line represents the cross section computed using
Eq.~(\ref{totcs-LO}), namely, the whole Born cross-section (the sum of
the Born cross-sections for all $J$'s) is multiplied by the enhancement
factor.
Each dashed line represents the sum of the cross sections for all $J$'s,
where only the $J=0$ cross section is multiplied by the enhancement
factor.
The red (solid and dashed) lines represent the cross sections for $gg
\to\ttb$ in the color-singlet channel, while the blue (solid and dashed)
line represent those in the color-octet channel.

The cross section in the singlet channel is more enhanced if we use the
overall prescription, Eq.~(\ref{totcs-LO}), since the force between the
color-singlet $\ttb$ pair is attractive and hence the enhancement factor
is larger than one.
The difference of the two prescriptions is sizable only above the $\ttb$
threshold and becomes maximal around $\mtt\simeq 400$~GeV,
where the difference is about 7\%.
On the other hand, the cross section in the octet channel is more
reduced if we use the overall prescription, since the force between the
color-octet $\ttb$ pair is repulsive and the enhancement factor is
(slightly) less than one.
The difference of the two prescriptions is at most 2\%.
The black lines (solid and dashed) represent the sum of the cross
sections for $gg\to\ttb$ in the above two channels.
The difference of the two prescriptions in this case is at most about
1\%, since the differences have opposite signs in the two channels and
are largely canceled.
Thus, the difference of the two prescriptions is rather small and much
smaller than other subleading corrections which we neglect in our analysis.
Furthermore, we have checked that the above tendencies are not changed
significantly by the ISR effects.
Therefore, for simplicity of our analysis, we will adopt the overall
prescription in the following analysis, namely, we will not decompose
the amplitude into different $J$'s.

In the case of $\qqb\to\ttb$, there is only the $J=1$ color-octet
channel at tree level.
Hence, the
enhancement factor multiplies the whole amplitude also in this case.

\subsection{Effects of large $\Gt$}\label{ss:topdecay}
\clfn

In this subsection we describe how we incorporate part of the
subleading corrections that are induced by the large top-quark width.
As an inevitable consequence of 
numerically integrating the fully differential
cross sections for $I\to \ttb \to \bwbw$, there is a significant
phase-space-suppression effect.
We partly compensate this effect,
which is related to a gauge cancellation inherent in the
inclusive cross section.

First, we briefly review existing theoretical studies on the
treatment of the top-quark width, in the cases with and 
without bound-state effects.
In the latter case, many schemes have been proposed
for incorporating
the top-quark width.
The use of the top-quark propagator in Eq.~(\ref{eqFWS}) is
called the fixed-width scheme (FWS).
It is widely used in simple analysis of the cross sections
which include the top quark as an unstable intermediate
particle. 
It is known, however, that subleading electroweak effects
are not properly treated in this scheme.
At present, the complex-mass scheme (CMS)~\cite{Denner:1999gp}
seems to be most
advanced from a practical point of view, 
due to the simplicity of its implementation.
In fact, for the process $e^+e^-\to W^+W^-$, which is
kinematically similar to $\ttb$ productions,
the fully differential cross section
has been computed incorporating the effects of 
$W$-boson width with NLO accuracy in this scheme, 
basically in all kinematical
regions~\cite{Denner:2005fg}.
For $\ttb$ productions in hadron collisions, 
the fully differential cross sections
is computed incorporating top-quark
width with LO accuracy in CMS, and various
differential cross sections
in different schemes were compared~\cite{Kauer:2002sn}.
In particular, the study has shown an agreement within errors between
all the calculated
cross sections in CMS and in FWS. 
(FWS is
simpler but less sophisticated than CMS.)

Regarding $\ttb$ productions in the threshold region
including the bound-state effects, studies on the finite-width
effects are most advanced in the total cross section for
$e^+e^-\to \ttb$.
The finite-width effects have been incorporated
with NNLO accuracy\footnote{
In the $\ttb$ threshold region, it is customary to count
$\Gt/\mt\sim{\cal O}(\alpha_W)\sim{\cal O}(\alpha_s^2)$.
}~\cite{Hoang:2004tg,Hoang:2010gu}, 
using the velocity-Non-Relativistic QCD
(vNRQCD) effective field theory framework~\cite{Luke:1999kz,Hoang:2002yy}.
Recently an NLO correction to the total cross section
arising from the single-top resonance
region has been pointed out and computed in \cite{Beneke:2010mp},
using unstable-particle effective field theory 
\cite{Beneke:2003xh,Beneke:2004km}.
On the other hand, in the corresponding
fully differential cross section the width effects are incorporated only
up to NLO accuracy~\cite{Peter:1997rk}
(apart from the contributions from the single-top resonance
region).
MC generators, developed specifically for simulation studies in the
threshold region of the $e^+e^-\to \ttb$ process, have incorporated both
bound-state effects and finite-width effects in
the LO approximation~\cite{Fujii:1993mk,Ikematsu:2003pg}.
For $\ttb$ productions in hadron collisions, only the $\ttb$
invariant-mass distributions have been computed with NLO accuracy,
incorporating the bound-state effects, finite-width effects
and ISR effects in the threshold region~\cite{Hagiwara:2008df,Kiyo:2008bv}.

One effect is known to be particularly important in
computing the fully differential cross sections
in the threshold region of $\ttb$ productions.
It is the phase-space-suppression
effect~\cite{Sumino:1992ai,Jezabek:1992np,Jezabek:1993rq}, which is
formally an NNLO effect of the top-quark width, but it seriously modifies the
shape of the sharply rising $S$-wave cross section
as a function of $\mtt$, after integrating the
differential cross section over the final-state phase-space
\cite{Sumino:1992ai}.
Let us briefly explain this effect.
The $\ttb$ cross section starts to rise below the
$\ttb$ threshold $\mtt=2\mt$ as a result of
formation of $\ttb$ resonances.
This means that the dominant kinematical configuration
is such that one of $t$ and $\tb$ is on-shell and
the other is off-shell.
Therefore, the phase-space of $bW$ which decayed from the
off-shell $t$ or $\tb$ is reduced as compared to the
on-shell case.
This suppresses the production cross section, and 
this effect is
automatically incorporated  if we integrate the LO
differential cross section numerically over the
phase-space of the final $\bwbw$.
A remarkable feature is that there is another effect
at NNLO which exactly cancels the phase-space-suppression effect, for
the integrated cross section 
at each $\ttb$ invariant-mass~\cite{Modritsch:1994hv,Kummer:1995id}. 
This is the Coulomb-enhancement
effect due to gluon exchanges
between $t$ and $\bb$ (decayed from $\tb$) and
between $\tb$ and $b$. 
The cancellation is guaranteed by gauge invariance (Ward
identity).
It protects the $\ttb$ resonance widths from being
determined by gauge-dependent off-shell width of the top quark.
Consequently the only surviving
NNLO effect to the $\ttb$ resonance widths
turns out to be the time-dilatation effect due to the relative
motion of $t$ and $\tb$ inside the resonances,
which is gauge independent.\footnote{
In principle, the momenta of $t$ and $\tb$ can be 
determined from the final state, in the LO approximation.
Hence, the relative motion is a gauge-independent quantity.
}

Thus, we face a problem when we compute differential cross sections
in the $\ttb$ threshold region
by a MC generator:
The phase-space-suppression effect is automatically incorporated,
while the Coulomb-enhancement
effect due to gluon exchanges
between $t$ and $\bb$ (or $\tb$ and $b$) 
is difficult to incorporate
in a MC generator.
(This is not yet achieved even in theoretical computations
of the $e^+e^-\to \ttb$ differential cross sections.)
Our prescription in this study is only effective.
Since we know that the phase-space-suppression effect is
canceled in the inclusive $\ttb$ cross section,
we multiply the $\ttb$ amplitude
$\cM^{(c)}_{\ttb}(I\to\ttb\to\bwbw)$ by an enhancement factor
such that the phase-space-suppression factor is
canceled.
In addition, we include the non-resonant diagrams,
which are formally ${\cal O}(\Gt/\mt)$ compared to the LO 
amplitude.

Hence, our full amplitude at the parton level
is given by
\bea
 \cM_{\bwbw}^{(c)}(I\to\bwbw) = 
\widetilde{\cM}_{\ttb}^{(c)} + \widetilde{\cM}^{(c)}_{\rm nr}
\label{formula-partonic}
\eea
with
\bea
&&
 \widetilde{\cM}^{(c)}_{\ttb}  = \cM^{(c)}_{\ttb} \times
 \left[\frac{\mt\Gt}{\sqrt{s_t}\Gt(s_t)}\cdot
 \frac{\mt\Gt}{\sqrt{s_{\tb}}\Gt(s_{\tb})}\right]^{\frac{1}{2}},
 \label{formula-partonic2a}
\\&&
 \widetilde{\cM}^{(c)}_{\rm nr} =
 \cM^{(c)}_{\ttb,{\rm tree}}\times
\left(1-\left[\frac{\mt\Gt}{\sqrt{s_t}\Gt(s_t)}\cdot
 \frac{\mt\Gt}{\sqrt{s_{\tb}}\Gt(s_{\tb})}\right]^{\frac{1}{2}}\right)
 + \cM_{\rm nr}^{(c)}\,,
 \label{formula-partonic2}
\eea
where
$\cM^{(c)}_{\ttb}$ is defined by Eqs.~(\ref{enhancement2}) and
(\ref{eq:ene}),
and $\cM^{(c)}_{\rm nr}$ denotes the sum of the 
tree-level non-resonant
diagrams.
The factor in the square bracket represents the inverse of
the phase-space-suppression factor
(see App.~B for the derivation),
where 
$\Gt(s_t)$ denotes the running top-quark width in unitary
gauge evaluated
at the top-quark invariant-mass $s_t$;
its explicit form is given in App.~B, Eq.~(\ref{Gt-unitarygauge}).
At large $\ttb$ invariant-masses,
the bound-state effects diminish, namely
${\cM}^{(c)}_{\ttb}\to\cM^{(c)}_{\ttb,{\rm tree}}$,
hence
the above $\bwbw$ amplitude is defined such that it
reduces to the tree-level amplitude,
$\cM_{\bwbw}^{(c)}\to\cM^{(c)}_{\ttb,{\rm tree}}+
\cM_{\rm nr}^{(c)}$.

At the differential level, the above
treatment of the cancellation of the phase-space-suppression
effect
is only effective, since 
the Coulomb-enhancement effect does not cancel the
phase-space-suppression effect at each kinematical point.
Nevertheless, we consider that a higher priority should
be given to the gauge cancellation mechanism that is inherent in the
inclusive cross section.
We also note that the replacement $E\to E'$ in 
the Green function, Eqs.~(\ref{enhancement2}) and
(\ref{eq:ene}), automatically incorporates
the time-dilatation effects to the resonance widths.\footnote{
The relation
$E'+i\Gt=\vec{p}^{\,2}/\mt$ (corresponding to
$\mt^2\to \mt^2-i\mt\Gt$) is relativistically correct,
so that the time-dilatation effect enters the lifetime of the
$\ttb$ system.
It can also be seen by the fact
$E'+i\Gt-\frac{\vec{p}^2}{\mt}\simeq E+i\Gt-\frac{\vec{p}^2}{\mt}
+\frac{1}{4\mt}\Bigl(\frac{\vec{p}^2}{\mt}-i\Gt\Bigr)^2
\simeq E+i\Gt\Bigl(1-\frac{\vec{p}^2}{2\mt^2}\Bigr)
-\frac{\vec{p}^2}{\mt}+\frac{\vec{p}^4}{4\mt^3}$,
where $\Bigl(1-\frac{\vec{p}^2}{2\mt^2}\Bigr)$ represents the
time-dilatation effect and $\frac{\vec{p}^4}{4\mt^3}$ represents a
relativistic correction.
}

\FIGURE[t]{
\includegraphics[width=.46\textwidth,clip]{part_ggqq.eps}
\quad
\includegraphics[width=.45\textwidth,clip]{part_thre.eps}
\caption{ 
Partonic
$\ttb$ invariant-mass
distributions for 
$gg\to\ttb$ and $\qqb\to\ttb$
in the individual channels.
Four lines in each figure are
computed from
(i) $|\cM^{(c)}_{\ttb}|^2$ (black dotted),
(ii) $|\widetilde{\cM}^{(c)}_{\ttb}|^2$ (blue dashed),
(iii) $|\cM^{(c)}_{\ttb}+\cM^{(c)}_{\rm nr}|^2$ (red dot-dashed),
and (iv) $|\widetilde{\cM}^{(c)}_{\ttb}+
\widetilde{\cM}^{(c)}_{\rm nr}|^2$ (green solid).
The figures in the right show the magnification of the
threshold region.
\label{fig:part_ggsng}}
}

In Fig.~\ref{fig:part_ggsng}, we compare the $\ttb$ invariant-mass
distributions for 
$gg\to\ttb$ and $\qqb\to\ttb$ in all the channels,
which are computed by integrating 
the following four cross sections over the $\bwbw$ phase-space, where
the $\ttb$ invariant-mass $\mtt$ is defined as the
invariant-mass of the final $\bwbw$ system:
(i) $|\cM^{(c)}_{\ttb}|^2$ (black dotted),
(ii) $|\widetilde{\cM}^{(c)}_{\ttb}|^2$ (blue dashed),
(iii) $|\cM^{(c)}_{\ttb}+\cM^{(c)}_{\rm nr}|^2$ (red dot-dashed),
and (iv) the absolute square of
our formula Eq.~(\ref{formula-partonic}) (green solid).
Comparing the distributions for
(i) and (ii), or, (iii) and (iv), we see that 
the phase-space-suppression effects are sizable especially
close to the threshold of
$gg\to\ttb$ in the color-singlet channel.
This is consistent with the explanation given above.
In particular, in each figure the difference between (iii) and (iv) is
hardly visible at large $\mtt$.
We may also compare the distributions for
(i) and (iii), or, (ii) and (iv), to see
the contributions of the non-resonant amplitude.
The contributions become comparatively larger at high energies for
gluon-fusion channels, 
since the contribution of the
$s$-channel diagram in $\cM^{(c)}_{\ttb}$ decreases,
while contributions of the single-resonant diagrams 
in $\cM^{(c)}_{\rm nr}$ increase.
On the other hand, for $\qqb$ channel, contributions from the
non-resonant diagrams are small everywhere.
\FIGURE[t]{
\begin{tabular}{cc}
\includegraphics[width=.45\textwidth,clip]{part_gg.eps}
&
\includegraphics[width=.45\textwidth,clip]{part_qq.eps}
\\
(a)&(b)
\end{tabular}
\caption{ 
$\ttb$ invariant-mass distributions for the processes $gg\to\bwbw$
 (color-summed) and $\qqb\to\bwbw$.
The red solid line is plotted using Eq.~(\ref{formula-partonic})
 for the partonic amplitude, while the black dashed line represents the
 Born-level cross section.
\label{fig:part_bwbw}}
}

In Fig.~\ref{fig:part_bwbw}(a) we plot the $\ttb$ invariant-mass
 distributions for $gg\to\bwbw$ (sum of the color-singlet
and octet channels),
using Eq.~(\ref{formula-partonic}) after integrating
over the $\bwbw$ phase-space (red solid line).
The $\ttb$ invariant-mass 
distribution at the Born level for $gg\to\bwbw$
in FWS is also plotted
(black dashed line).
The difference of the two lines signifies
the bound-state effects, after including the
non-resonant diagrams and compensating the
phase-space-suppression effect.
The enhancement of the distribution by the bound-state
effect is visible not only in the threshold region but also up to
about $\mtt=500$~GeV.
The enhancement factors are about 1.05 and 1.02 at
$\mtt=400$~GeV and 500~GeV, respectively.
The $\ttb$ invariant-mass 
distributions for $\qqb\to\bwbw$ are also plotted
in Fig.~\ref{fig:part_bwbw}(b).
The enhancement factor is smaller than unity because
of the repulsive force, whose values are 
about 0.96 and 0.98
at $\mtt=400$~GeV and 500~GeV, respectively.

\subsection{Inclusion of ISR effects and $K$-factors}
\label{sec:isr}

In this subsection we explain how we incorporate the ISR
effects in the cross sections in our framework.
In addition, we determine the $K$-factors to match our
predictions for the $\ttb$ invariant-mass distributions to
the available NLO predictions.

In hadron collisions, it is important to include the ISR effects.
In our framework, they are incorporated by connecting
the differential cross sections computed from the matrix elements
Eq.~(\ref{enhancement2}) to a parton-shower simulator such as {\tt
PYTHIA}~\cite{Sjostrand:2006za} or {\tt HERWIG}~\cite{Corcella:2000bw}.
In addition, we include ``$K$-factors'' as the normalization constants
of the cross sections for $I\to\bwbw$ in the individual
channels.\footnote{%
Note that the parton shower simulators incorporate ISR effects
by way of stochastic
processes, and that the $\ttb$ invariant-mass 
distributions handed to the simulators at the parton level
are not affected by the ISR effects.
}
The $K$-factors are determined such that the $\ttb$ invariant-mass
distribution for each channel 
in the threshold region matches the corresponding NLO prediction
in the threshold region.
We also extrapolate these $K$-factors to the large $\mtt$ region.
The main reason to do so is a lack of the NLO 
predictions in the large $\mtt$ region 
for the individual channels:
The present theoretical prediction for
the NLO $\ttb$ invariant-mass
distribution in the large $\mtt$ region
is provided numerically
only for the color-summed cross section
for on-shell top-quark productions
\cite{Mangano:1991jk}.
Theoretically, by naive extrapolation of the
$K$-factors, we reproduce the double-logarithmic terms
of the cross sections correctly in the large $\mtt$ region,
due to the universal structure of soft-gluon emissions;
on the other hand, we do not reproduce the
single-logarithmic and non-logarithmic terms.

\FIGURE[t]{
\includegraphics[width=0.6\textwidth,clip]{kfac_thr14.eps}
\caption{
 $\mtt$ dependence of the $K$-factors for $\ttb$ production in $gg$
 color-singlet (top), color-octet (middle) and $\qqb$ (bottom) channels
 at the LHC $\sqrt{s}=14$~TeV.
 The dashed, solid and dotted lines are obtained with $\mu_R=\mu_F=
 \kappa \mt$ with $\kappa=0.5$, 1 and 2, respectively.
 \label{fig:kfac}}
}

The NLO corrections to the
$\ttb$ invariant-mass distributions in the threshold region
are known for the individual channels 
\cite{Hagiwara:2008df,Kiyo:2008bv}.
The corrections are given in terms of 
the hard-correction factors and the gluon radiation
functions.
The major difference of the
predictions of \cite{Hagiwara:2008df} and \cite{Kiyo:2008bv} is
that in the latter predictions contributions from
high $\sqrt{\hat{s}}$ (the c.m.\ collision energy of the
initial partons) are included more accurately.\footnote{
In the gluon radiation functions, the terms enhanced by 
plus-distributions or delta-functions 
as $z\to 1$ are common in  
\cite{Hagiwara:2008df} and \cite{Kiyo:2008bv}, while
non-enhanced terms differ.
} 
Hence, we use the latter predictions to
compute the $K$-factors\footnote{
In \cite{Kiyo:2008bv}, effects of resummation of
threshold logs are also
examined and found to enhance the normalization at 10\% level.
See also \cite{Beneke:2009rj,Beneke:2009ye,Ahrens:2010zv}.
}.
We can determine the $K$-factors by taking the ratios of these NLO  
partonic cross sections and 
our (LO) partonic cross sections given in 
Sec.~\ref{ss:topdecay}.
Since the NLO cross sections~\cite{Hagiwara:2008df,Kiyo:2008bv} 
do not include contributions from non-resonant diagrams
${\cal M}_{\rm nr}^{(c)}$, accordingly we incorporate only
the contributions from the
resonant diagrams $\widetilde{\cal M}_{\ttb}^{(c)}$
in the LO cross sections 
when we compute the $K$-factors.
(In most of the threshold region, the effect of
$\widetilde{\cal M}_{\rm nr}^{(c)}$ is irrelevant in any case, since
resonant diagrams dominate.)
Furthermore, in calculating the $K$-factors, we use the CTEQ6M
PDFs~\cite{Pumplin:2002vw} and the 2-loop running of 
the strong coupling constant $\alpha_s$
for the NLO $\mtt$ distribution, while
the CTEQ6L1 PDFs and the 1-loop running of $\alpha_s$ are used
for the LO distribution.
(We find that the $K$-factors obtained by using the MSTW2008
PDFs~\cite{Martin:2009iq} are quite similar.)

In general, the $K$-factors depend on $\mtt$.
We first examine $\mtt$-dependences of the $K$-factors as we
choose different renormalization and factorization
scales, $\mu_R$ and $\mu_F$.
The renormalization 
scale $\mu_R$ enters the NLO
formula as the scale of the strong coupling constant and also through the
logarithmic term in the hard-vertex function; see 
Eq.~(3.2) of \cite{Kiyo:2008bv}.
On the other hand, the factorization scale $\mu_F$ enters 
the NLO formula as the scale of
the parton distribution functions (PDFs) and through the terms with
$\ln\left(\mtt^2/\mu_F^2\right)$ in the gluon radiation functions; see
Eqs.~(3.4-3.7) of \cite{Kiyo:2008bv}.
We find that, the $\mtt$-dependences of the $K$-factors can be
relatively flat in the threshold region, with appropriate choices of
$\mu_R$ and $\mu_F$.
In this case, extrapolation of the $K$-factors from the threshold region
to the high $\mtt$ region can be performed trivially.
Indeed, for simplicity of our analysis, we take the $K$-factors to be
independent of $\mtt$.
In Fig.~\ref{fig:kfac}, we plot the $K$-factors of the $\ttb$
invariant-mass distributions
in the individual channels at the LHC with $\sqrt{s}=14$~TeV, in
the cases that we choose the scales as
$\mu_R=\mu_F=\kappa \mt$
with $\kappa=0.5,1,2$.
As can be seen, the $\mtt$-dependence of the
$K$-factors are mild.
We have also examined the $K$-factors corresponding to
the LHC with $\sqrt{s}=7$~TeV and
Tevatron with $\sqrt{s}=1.96$~TeV; we find that
the $K$-factors are only mildly dependent on
$\mtt$ also in these cases.
In Table~\ref{tab:kfac}, we list the numerical values of the
$K$-factors for all the channels corresponding to the
LHC $\sqrt{s}=14$~TeV, 7~TeV and Tevatron
$\sqrt{s}=1.96$~TeV, obtained at $\mtt=2\mt$ for
$\mu_R=\mu_F=\kappa \mt$ with $\kappa=0.5$, 1, 2.
The values of the $K$-factors for $\kappa=1$ will be
used in the following.

\TABLE[bt]{
\begin{tabular}{|c||c|c|c||c|c|c||c|c|c|}
 \multicolumn{1}{c}{} &
 \multicolumn{3}{c}{LHC 14~TeV} &
 \multicolumn{3}{c}{LHC  7~TeV} &
 \multicolumn{3}{c}{Tevatron} \\
 \hline
 $\kappa$
 & $gg^{[c=1]}$ & $gg^{[c=8]}$& $\qqb$
 & $gg^{[c=1]}$ & $gg^{[c=8]}$& $\qqb$
 & $gg^{[c=1]}$ & $gg^{[c=8]}$& $\qqb$ \\
 \hline
 0.5 & 0.79& 1.02& 0.88& 0.87& 1.13& 0.89& 1.30& 1.72& 0.87 \\
 1   & 1.14& 1.39& 1.16& 1.31& 1.60& 1.18& 2.11& 2.60& 1.18 \\
 2   & 1.48& 1.75& 1.42& 1.75& 2.07& 1.45& 2.95& 3.48& 1.45 \\
 \hline
\end{tabular}
\caption{
 $K$-factor normalization constant for each channel ($gg$ color-singlet,
 octet and $\qqb$) for the LHC $\sqrt{s}=14$~TeV, 7~TeV and the
 Tevatron, with setting the factorization and renormalization scales to
 $\mu_R=\mu_F=\kappa\mt$ with $\kappa=0.5$, 1, 2.
 \label{tab:kfac}}
}
\FIGURE[t]{
\includegraphics[width=0.6\textwidth,clip]{fullnlo.eps}
\caption{
Ratio of the two color-summed $\ttb$ invariant-mass distributions: The
 former is our prediction including the $K$-factors but omitting the
 non-resonant diagrams, and the latter is the NLO prediction by {\tt
 MC@NLO} for on-shell top quarks.
 \label{fig:ratio-mc@nlo}}
}

We check consistency of our $K$-factor normalization in the large
$\mtt$ region, by comparing our prediction for the color-summed $\ttb$
invariant-mass distribution with the NLO prediction.
In Fig.~\ref{fig:ratio-mc@nlo} we plot the ratio of these two cross
sections.
In the former distribution, we
include the $K$-factors, while we do not include
the non-resonant diagrams
$\widetilde{\cal M}_{\rm nr}^{(c)}$
[c.f.\ (ii) of Fig.~\ref{fig:part_ggsng}].
The latter
distribution is computed for
the on-shell $\ttb$ productions, by {\tt
MC@NLO}~\cite{Frixione:2002ik,Frixione:2003ei} with CTEQ6M PDFs with 
a scale choice $\mu_F=\mu_R=\sqrt{\mt^2+p_{T,t}^2}$.
As can be seen, both cross sections are 
mutually consistent within
2\% accuracy up to $\mtt=800$~GeV.

Including non-resonant diagrams in a way
that the gauge cancellation holds effectively, 
our final formula for the 
matrix element at parton level reads
\bea
 \cM_{\bwbw}^{(c)}(I\to\bwbw) = \sqrt{K}\,
\left[
\widetilde{\cM}_{\ttb}^{(c)} + \widetilde{\cM}^{(c)}_{\rm nr}
\right] .
\label{formula-partonic-final}
\eea
$\widetilde{\cM}_{\ttb}^{(c)}$ and
$\widetilde{\cM}^{(c)}_{\rm nr}$ are given in 
Eqs.~(\ref{formula-partonic2a}) and (\ref{formula-partonic2}).

\section{Event Generation and Top-Quark Distributions}
\clfn

In this section, we present numerical evaluations 
of various kinematical distributions of the top-quark
computed from the $pp\to\bwbw$ cross section,
using the theoretical framework explained in the
previous section.
In particular we study the bound-state effects on these
distributions.

Our numerical calculations are carried out based on the {\tt MadGraph}
output~\cite{Stelzer:1994ta} which makes use of the {\tt HELAS}
subroutines~\cite{Murayama:1992gi} for helicity-amplitude calculations.
The original {\tt MadGraph} output code has been modified to implement
the color-decomposition and to include the bound-state effects via the
Green functions.
For the convenience of the readers, we collect the formulas necessary for
decomposing amplitudes into the color-singlet and octet components of
the $\ttb$ (or $b\bb$) system in Appendix~\ref{app:color}.
In particular, we discuss how to implement the color decomposition into
the {\tt MadGraph} notation.
The bound-state correction factor
\begin{align}
 \cG^{(c)}(E,\vec{p}) = \frac{G^{(c)}(E,\vec{p})}{G_0(E,\vec{p})},
 \label{eq:corr}
\end{align}
c.f.\ Eq.~(\ref{enhancement2}),
is pre-tabulated to save time for computing the momentum-space
Green functions\footnote{%
The $S$-wave Green function depends only on $|\vec{p}|$ but not on
the direction of the three-momenta.}.

We perform phase-space integrations using {\tt
BASES/SPRING}~\cite{Kawabata:1985yt}, or alternatively by adapting our
code to {\tt MadEvent}~\cite{Maltoni:2002qb,Alwall:2007st} utilities
(ver.\ 4.4.42),
where both tools are able to generate unweighted
events at the partonic final-state level.
For each event, we assign the specific color-flow according to an
ordinary manner, except the color-singlet channel, as explained in
Appendix~\ref{app:color}.
The generated events can be subsequently provided e.g.\ to {\tt
PYTHIA} for simulations of parton-showering and hadronizations.

In the main body of this paper, we do not consider the
decay of $W$'s but consider only the observables constructed
from the $\bwbw$ final state.
The $W$-boson decays can be incorporated at the {\tt PYTHIA} stage,
where however the polarization of $W$-bosons cannot be taken into
account.
Alternatively, one can calculate the helicity amplitudes including the
decay of $W$-bosons by specifying a decay mode for each
$W$-boson.
In Appendix~\ref{app:wdecay}, as a sample case, we examine the
distributions of dileptons in the dilepton mode, where both
$W$'s decay leptonically, and study the effects of $W$-boson
polarization and bound-state corrections.

Below we show the results at the partonic
$\bwbw$ final-state level.
We do not discuss the parton-showering and
hadronization effects, 
in order to concentrate on the examination of bound-state
effects.
For the parton distribution functions, we use the CTEQ6L1
parameterization with the LO evolution (1-loop running)
of the QCD coupling
constant.
We set the renormalization and factorization scales to
$\mu_{R}=\mu_{F}=\mt$ and incorporate the $K$-factors obtained in
Sec.~\ref{sec:isr} to the cross sections in
the individual channels. 
[The final formula for the matrix element is given by 
Eq.~(\ref{formula-partonic-final}).]
We set the top-quark pole-mass, the (tree-level)
on-shell top-quark width and the
strong coupling constant as
$\mt=173$~GeV, $\Gt=1.49$~GeV
and 
$\alpha_s(M_z)=0.1298$,
respectively.

\FIGURE[t]{
\includegraphics[width=.45\textwidth,clip]{dsdmtt_lhc.eps}\quad
\includegraphics[width=.45\textwidth,clip]{dsdmtt_thre.eps}
\\
(a)\hspace{70mm}(b)
\caption{$\ttb$ invariant-mass distribution in $pp\to\bwbw$ 
 at $\sqrt{s}=14$~TeV.
 Green solid line is our full prediction and blue dashed line is the
 Born-level prediction.
 The NLO $\ttb$ production computed by {\tt MC@NLO} is also
 plotted in red dots.
 Right figure is the magnification of the threshold region.
 \label{fig:mtt}}
}
In Fig.~\ref{fig:mtt}(a), we plot the $\ttb$ invariant-mass
distribution in $pp\to\bwbw$ production at $\sqrt{s}=14$~TeV.
The $\ttb$ invariant-mass $\mtt$ is defined as the
invariant-mass of the final $\bwbw$ system.
The green solid line represents the full result which includes the bound-state
effects as well as the $K$-factors, and the blue dashed
line represents the Born-level result (the LO prediction in the
conventional perturbative QCD approach).
Fig.~\ref{fig:mtt}(b) shows a magnification of the same cross
sections in the
threshold region.
As shown in \cite{Hagiwara:2008df,Kiyo:2008bv}, 
theoretically the bound-state effects can be seen most clearly
in the shape of the $\mtt$ distribution in the threshold region.
One can see that the cross section is enhanced over the
Born cross-section
significantly by the bound-state effects, 
and there appears a broad peak
below the threshold corresponding to the $^{1}S_{0}$ resonance state
in the color-singlet $\ttb$ channel.
Far above the threshold, the bound-state effects 
disappear and the cross section approaches
the Born-level distributions, up to the $K$-factor normalization.

In the same figures, we also
compare our prediction 
with the NLO $\mtt$ distribution computed by 
{\tt MC@NLO}~\cite{Frixione:2002ik,Frixione:2003ei} with 
CTEQ6M PDFs and the
scale choice of $\mu_F=\mu_R=\sqrt{\mt^2+p_{T,t}^2}$.
The latter prediction includes the full NLO QCD corrections
(but not the Coulomb resummation) for the on-shell $\ttb$ productions;
we switched on an option of {\tt MC@NLO} to incorporate
off-shellness of the top-quarks effectively
by re-weighting the cross section by skewed Breit-Wigner
functions~\cite{Frixione:2007zp}, so that the cross section is non-zero
below the threshold.
(However, non-resonant diagrams are not incorporated.)
Below and near the threshold, our prediction is
much larger than the {\tt MC@NLO} prediction,
due to the bound-state formation.
The two cross sections become approximately equal from around
$\mtt\sim370$--380~GeV up to larger $\mtt$.
Note that, in Fig.~\ref{fig:ratio-mc@nlo}, the contributions
from non-resonant diagrams are not included in our full 
prediction, whereas in Figs.~\ref{fig:mtt} they are included.
Integrating the distributions over $\mtt$,
the total cross section by our full (Born-level) calculation
is estimated as
$\sigma_{\bwbw}=855$~pb (633~pb), while we obtain
$\sigma_{\ttb}=816$~pb as the {\tt MC@NLO} prediction.

\FIGURE[t]{
\includegraphics[width=.45\textwidth,clip]{dsdmtt_lhc07.eps}\quad
\includegraphics[width=.45\textwidth,clip]{dsdmtt_thre07.eps}
\caption{The same as Fig.~\ref{fig:mtt}, but for the LHC
 $\sqrt{s}=7$~TeV. \label{fig:mtt07}}
}
The shape of the $\mtt$ distribution at the LHC 7~TeV, shown in
Fig.~\ref{fig:mtt07}, is similar to that for the LHC 14~TeV.
The
total cross sections are estimated to be
158~pb, 106~pb and 146~pb by our
full, Born-level calculations and {\tt MC@NLO}, respectively. 
\medbreak

Let us examine other distributions of the top quark.
From Figs.~\ref{fig:mtt} and \ref{fig:mtt07}, 
it is obvious that the phase-space region,
where the bound-state effects are important,
corresponds to a rather limited portion of the full top-quark
events produced at the LHC.
Thus, in various distributions formed by the full 
events, the bound-state
effects may well
be negligible in practice.
In order to examine the bound-state effects closely,
in the following we consider the events restricted
by $\mtt\leq370$~GeV (except where otherwise stated),
instead of considering the full events.
They amount to about 9\% (8\%) of the full events 
according to our calculation
with (without) the bound-state corrections at  
the LHC $14$~TeV.
Due to the large $\ttb$ cross sections at the LHC, still
a large number of such near-threshold events
would be accumulated.
In this paper, we do not discuss the important
subject of how to measure
$\mtt$ in real experiments, which 
requires detailed studies of errors and fake
solutions;
one may find them in earlier studies 
\cite{Frederix:2007gi,Mahlon:2010gw}. 

\FIGURE[t]{
\includegraphics[width=.9\textwidth,clip]{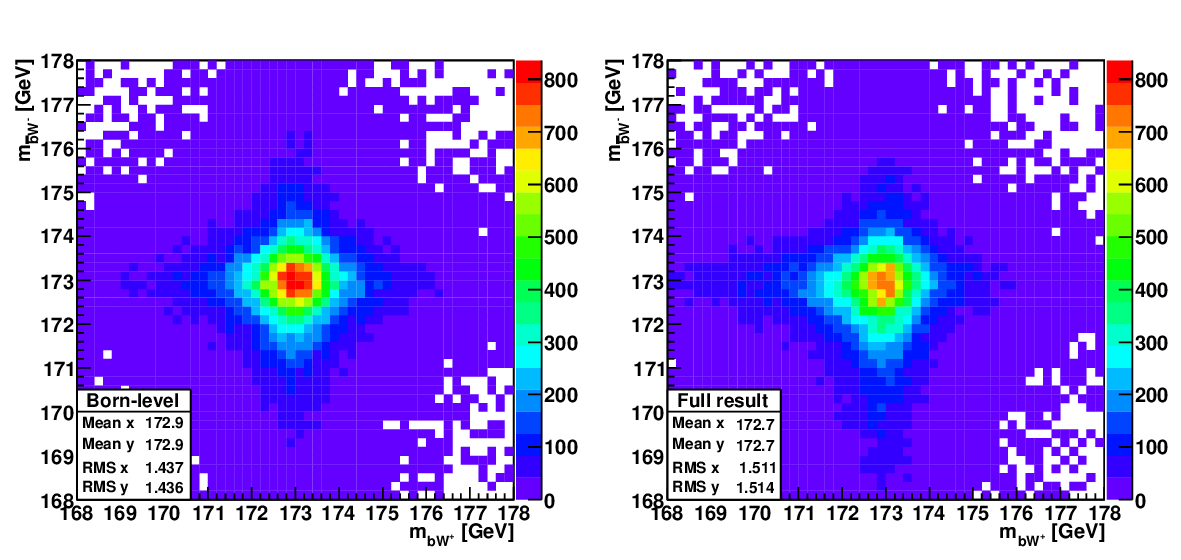}\\
(a)\hspace{70mm}(b)
\caption{Two-dimensional density plots of the $bW^{+}$ and
 $\bb W^{-}$ invariant-masses, for the events with $\mtt\leq370$~GeV at
 the LHC $\sqrt{s}=14$~TeV.
 Left figure (a) is the Born-level prediction and right figure 
(b) is our full
 result (including the bound-state corrections and $K$-factors).
 The mean value and the root-mean-square value displayed in each
 figure are calculated for the events within the frame of the figure.
 \label{fig:mt1mt2}}
}
One observes
a characteristic bound-state effect 
in the $(bW^{+})$-$(\bar{b}W^{-})$ double-invariant-mass distribution.
In Figs.~\ref{fig:mt1mt2}, we show the
density plots of the invariant-masses of the 
$bW^{+}$ and $\bar{b}W^{-}$ systems, given by
(a) the Born-level prediction and
(b) our full prediction.
In each figure,
the number of events is normalized to 100,000 in total, and the
number of events per bin (0.2~GeV$\times$0.2~GeV) is plotted with 
graded colors.
The Born-level prediction (a)
is essentially determined by the product
of the Breit-Wigner functions,
hence the distribution is almost reflection
symmetric with respect to 
the on-shell lines $(p_b+p_{W^+})^2=\mt^2$
and $(p_{\bar{b}}+p_{W^-})^2=\mt^2$.
By contrast, the distribution by our full prediction (b)
is not symmetric and
biased towards the configuration, where one
of $t$ or $\bar{t}$ is on-shell and the other has
an invariant-mass smaller than $\mt$.
In fact, such a configuration is known to be the
dominant configuration just below the threshold 
in $e^+e^-\to\ttb$~\cite{Sumino:1992ai,Ikematsu:2003pg}, although in that case 
deviation from
the double Breit-Wigner distribution [Fig.~\ref{fig:mt1mt2}(a)]
is more prominent.
(Note that, below the threshold, $t$ and $\bar{t}$
cannot become simultaneously on-shell.)

In order to quantify the correlated deformation of
the double-invariant-mass distribution,
we count the fraction of the events for which 
both or either of the
$bW^{+}$ and $\bar{b}W^{-}$ invariant-masses satisfy
\begin{align}
 |m_{bW}-\mt|\leq k\Gt, 
\label{eq:kGt} 
\end{align}
where $k=1,2,\dots,5$.
These fractions are tabulated in Table~\ref{tab1} for
our full prediction and for the Born-level prediction.
The bound-state effect reduces the fraction for which
both invariant-masses are close to on-shell more
than the fraction for which either of the invariant-masses 
is close to on-shell.
In the former case, the change of the fraction by the bound-state
effect amounts up to about 4\%.
For comparison, we also tabulate the same
fractions for the full
events; in this case,
the variation of the fractions are small
and at most 1\%.
In any case, a proper understanding of this effect would
be important, since it potentially biases the
mass cut and may affect, for instance,
the top-quark mass measurement with high accuracy.

\TABLE[t]{
\begin{tabular}{|c|cc||cc|}
 \multicolumn{1}{c}{} &
 \multicolumn{2}{c}{$\mtt\leq370$~GeV} &
 \multicolumn{2}{c}{Full events} \\
 \hline
 $k$ & both [\%]& either [\%]& both [\%]& either [\%] \\
 \hline
 1 & 37.0 (41.1) & 87.0 (87.6) & 46.8 (47.7) & 90.3 (90.4) \\
 2 & 55.5 (59.0) & 95.2 (95.2) & 67.6 (68.3) & 97.0 (97.0) \\
 3 & 64.2 (66.3) & 97.3 (97.1) & 76.3 (76.8) & 98.5 (98.5) \\
 4 & 69.2 (70.3) & 98.1 (98.0) & 81.0 (81.4) & 99.1 (99.1) \\
 5 & 72.3 (72.9) & 98.6 (98.4) & 83.9 (84.2) & 99.4 (99.4) \\
 \hline
\end{tabular}
\caption{A fraction of events which satisfy $|m_{bW}-\mt|\leq k\Gt$ for
 both or either of the $bW$ invariant-masses.
 The events with $\mtt\leq370$~GeV as well as the full events at the LHC
 14~TeV are considered.
 In the bracket is shown the result in Born-level.
\label{tab1}}
}

Let us explain the mechanism how the bound-state effects
alter the double-invariant-mass distribution.
As shown in Appendix A,
the leading part of the $\ttb$ amplitude $\cM^{(c)}_{\ttb}$ 
has a form
\bea
&&
\cM^{(c)}_{\ttb}\propto
\left\langle
\frac{1}{\mtt-[2\sqrt{\hat{p}^2+\mt^2}+V_{\rm QCD}^{(c)}(r)] }
\right\rangle\,
\nn\\
&&~~~~~~~~~~~~~~~~~
\times
\left(
\frac{1}{\hat{p}_t^0-\sqrt{\hat{p}^2+\mt^2}}+
\frac{1}{\hat{p}_{\bar{t}}^0-\sqrt{\hat{p}^2+\mt^2}}
\right)\,
\label{symbolic-expl}
\eea
where $\hat{p}_t^\mu\equiv \hat{p}_b^\mu+\hat{p}_{W^+}^\mu$ and 
$\hat{p}_{\bar{t}}^\mu\equiv\hat{p}_{\bar{b}}^\mu+\hat{p}_{W^-}^\mu$
are defined in the $\ttb$ c.m.\ frame, and
$\hat{p}\equiv |\vec{\hat{p}}_t|$ denotes the magnitude of the
top-quark three-momentum in this frame.
The first factor on the right-hand side
$\langle (\mtt-H)^{-1}\rangle$, with
$H=2\sqrt{\hat{p}^2+\mt^2}+V_{\rm QCD}^{(c)}(r)$ 
[c.f.\ Eq.~(\ref{H-naturalchoice})],
denotes the Green function of the $\ttb$ system.
We suppress the top-quark width for simplicity.
In the case that $\ttb$ is in the singlet channel
($c=1$),
the potential energy between $t$ and $\bar{t}$ is negative,
$V_{\rm QCD}^{(1)}(r)<0$.
Therefore, the denominator of the Green function 
become close to zero (hence, the Green function
is most enhanced) if 
$\hat{p}$ is somewhat larger than the on-shell momentum
$p_{\rm OS}\equiv \sqrt{{\mtt^2}/{4}-\mt^2}$,
i.e.,
$\hat{p}> p_{\rm OS}$.
On the other hand, the second factor
$({p_t^0-\sqrt{\hat{p}^2+\mt^2}})^{-1}+
({p_{\bar{t}}^0-\sqrt{\hat{p}^2+\mt^2}})^{-1}$ is most
enhanced when $\hat{p}=p_{\rm OS}$, since
$p_t^0+p_{\bar{t}}^0=\mtt$.
Thus, there is a competition between the two factors
on the right-hand side of Eq.~(\ref{symbolic-expl}).
As a consequence, the dominant configuration is the one
in which neither of the two factors
are maximal.
In fact, in the dominant configuration one of $t$ and $\tb$
is on-shell and the other is off-shell:
$\hat{p}_t^2=\mt^2$, $\hat{p}_{\bar{t}}^2<\mt^2$, or,
$\hat{p}_t^2<\mt^2$, $\hat{p}_{\bar{t}}^2=\mt^2$.
The effect is opposite in the case that $\ttb$ is in the 
octet channel ($c=8$).
Since the magnitude of the octet
potential is much smaller than the singlet potential, 
$V_{\rm QCD}^{(8)}/V_{\rm QCD}^{(1)}=-1/8$,
the bound-state effect turns out to
be much larger  in the singlet
channel than in
the octet channel.

\FIGURE[t]{
\includegraphics[width=0.45\textwidth,clip]{phat_370_sng.eps}\quad
\includegraphics[width=0.45\textwidth,clip]{phat_370_oct.eps}
\\
~~~~(a)\hspace{65mm}~~~(b)
\caption{
Top-quark momentum ($\hat{p}\equiv |\vec{\hat{p}}_t|$) distributions
in the partonic center-of-mass frame
 for a fixed $\mtt=370$~GeV,
 (a) for the color-singlet channel, and (b) for the octet channel.
The solid lines represent our full prediction, after omitting
the non-resonant diagrams.
 The dotted lines represent corresponding distributions at the Born-level.
 In Fig.~(a), the distribution by the non-relativistic formula
(replacing $E'\to E$ in the Green
 function) is also plotted (the red dashed line).
 \label{fig:momdist}}
}
Displayed in Figs.~\ref{fig:momdist}(a) and (b) are
the top-quark momentum ($\hat{p}$) distributions
of the events with $\mtt=370$~GeV
({\it not} with $\mtt\leq 370$~GeV), for the color-singlet and
octet channels, respectively. 
To see essential features,
only the $\ttb$ diagrams are taken into
account and the $K$-factors are not included.
In each figure,
the black solid (blue dotted) line shows our full prediction (Born-level 
prediction).
The peak momentum for each distribution is shown with a
vertical line.
The peak momenta of the Born-level distributions
are (to a good approximation)
the on-shell momentum, $\hat{p}_{\rm peak}\approx
p_{\rm OS}
=65.5$~GeV.
We see that the bound-state effects shift
the peak momentum by
about 0.7~GeV to a
larger value for the color-singlet distribution, while
the peak momentum 
of the color-octet channel is shifted only by 
{50~MeV} to a smaller
value.
In the color-summed cross section, the peak momentum
is shifted to a larger value.
Consequently,
one of the invariant-masses of the 
$bW^{+}$ and $\bar{b}W^{-}$ systems is reduced below $\mt$.
The integral of this effect over 
the region $\mtt\leq 370$~GeV can be seen in
Fig.~\ref{fig:mt1mt2}(b).

One may suspect that the above shifts of the invariant-masses (or the
shift of the peak momentum) may be an artifact of our specific method
to interpolate the $\ttb$ cross sections in the threshold region and 
in the higher $\mtt$ region.
To check this,
let us estimate the size of the shift of the peak momentum
at $\mtt=370$~GeV and compare it
with the above prediction.
The distance a top-quark propagates before it decays is
estimated as
$\gamma c \tau=\mtt/(2\mt\Gt)=0.72~\mbox{GeV}^{-1}
=1/(1.4~\mbox{GeV})$.
This distance is considered to be within the range where
the potential $V_{\rm QCD}^{(1)}(r)$ can be
estimated perturbatively, although the 1-loop potential
tends to underestimate the bound-state effect.\footnote{
See e.g. \cite{Sumino:2005cq} for the recent status of the
perturbative prediction for the QCD potential.
}
The shift of the average momentum may be estimated by
$\bigl[\{ \mtt-V_{\rm QCD}^{(1)}(\gamma c \tau)\}^2/4
-\mt^2\bigr]^{1/2}
-p_{\rm OS}=0.66$~GeV.
Hence, the effect seen in Fig.~\ref{fig:momdist}(a) seems to be physical.

We note that the effect elucidated here is a kind of
effect that can never be seen in 
perturbative QCD computations for the {\it on-shell} 
$\ttb$ productions,
such as those given in 
\cite{Frixione:2002ik,Frixione:2003ei,Frixione:2007nw}.
This is because, the effect 
originates from the exchange of Coulomb gluons between
off-shell $t$ and on-shell $\bar{t}$ (or {\it vice versa}).
Our full prediction correctly incorporates the 
(gauge-independent) LO off-shellness
of the top quark as dictated by the exchange of Coulomb gluons,
which is crucial for predicting the deformations of the
top-quark momentum distribution and the double-invariant-mass 
distribution of the 
$bW^{+}$ and $\bar{b}W^{-}$ systems.

Now we are in a position to understand the origin of 
the abnormally large enhancement
of the cross section, which we observed in 
Sec.~2.1, in the case that we use the non-relativistic formula
for the differential cross section at large $\mtt$;
see the red dot-dashed line in Fig.~\ref{fig:sigh_ene}.
The non-relativistic formula corresponds to replacing the
Hamiltonian $H=2\sqrt{\hat{p}^2+\mt^2}+V_{\rm QCD}^{(c)}(r)$ by
$H=2\mt +{\hat{p}^2}/{\mt}+V_{\rm QCD}^{(c)}(r)$
in the Green function in Eq.~(\ref{symbolic-expl}).
Thus, the non-relativistic formula overestimates the kinetic 
energy of the $\ttb$ system in the large $\mtt$ region,
$2\mt +{\hat{p}^2}/{\mt}>2\sqrt{\hat{p}^2+\mt^2}$.
For this reason, the two factors on the right-hand side of 
Eq.~(\ref{symbolic-expl}) can be brought close to maximal 
simultaneously  with a nearly on-shell momentum,
$\hat{p}\simeq p_{\rm OS}$,
since all the denominators in this expression nearly
vanish.
Since the individual factors are made of pole-type functions,
applying
an inaccurate kinematical relation only in one of the denominators
can lead to a substantial overestimate of  the cross section.
In Fig.~\ref{fig:momdist}(a) we also plot
the top-quark-momentum distribution computed with
the non-relativistic formula, Eq.~(\ref{enhancement2}) after
the replacement $E'\to E$ for the singlet channel
(red dashed line). 
As can be seen, the peak momentum approaches the
on-shell momentum and the distribution is more enhanced
around the peak, compared to our full prediction.
\\
\FIGURE[t]{
\includegraphics[width=.45\textwidth,clip]{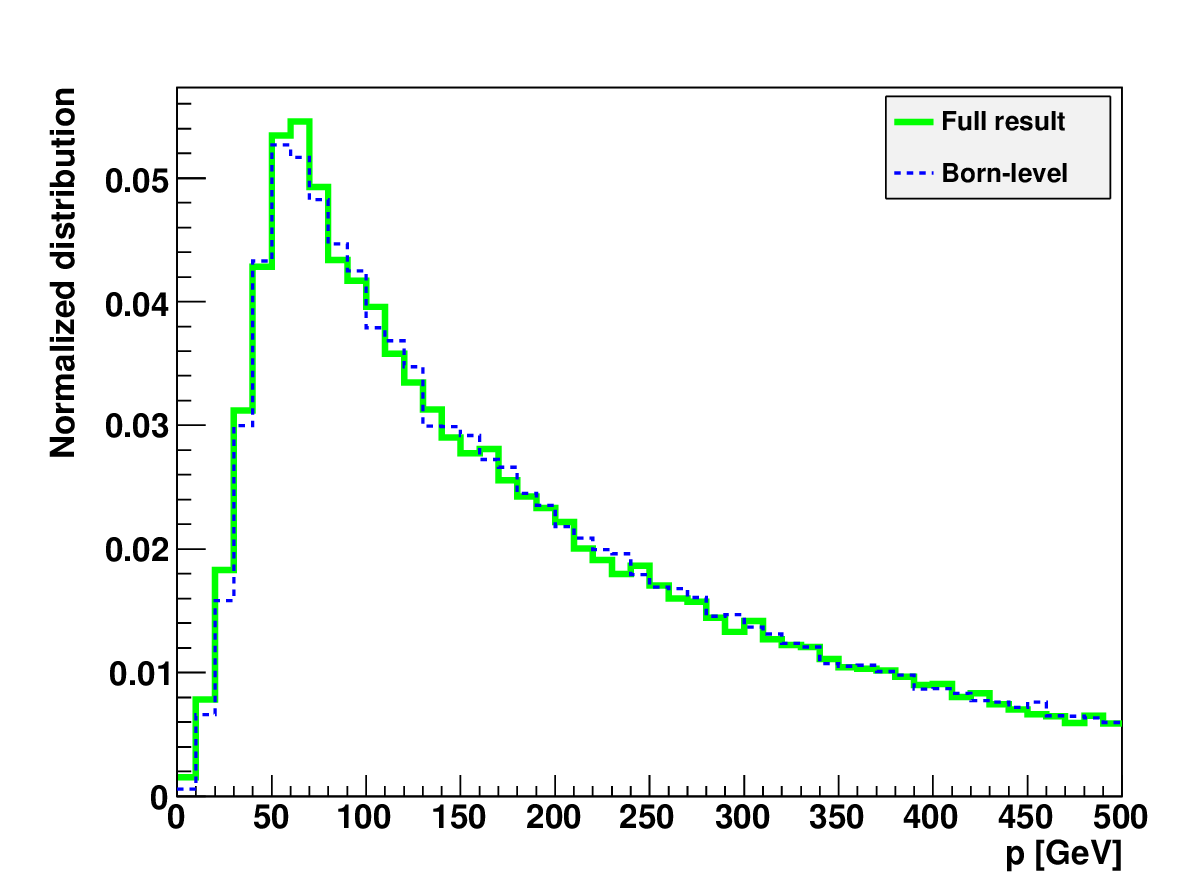}\quad
\includegraphics[width=.45\textwidth,clip]{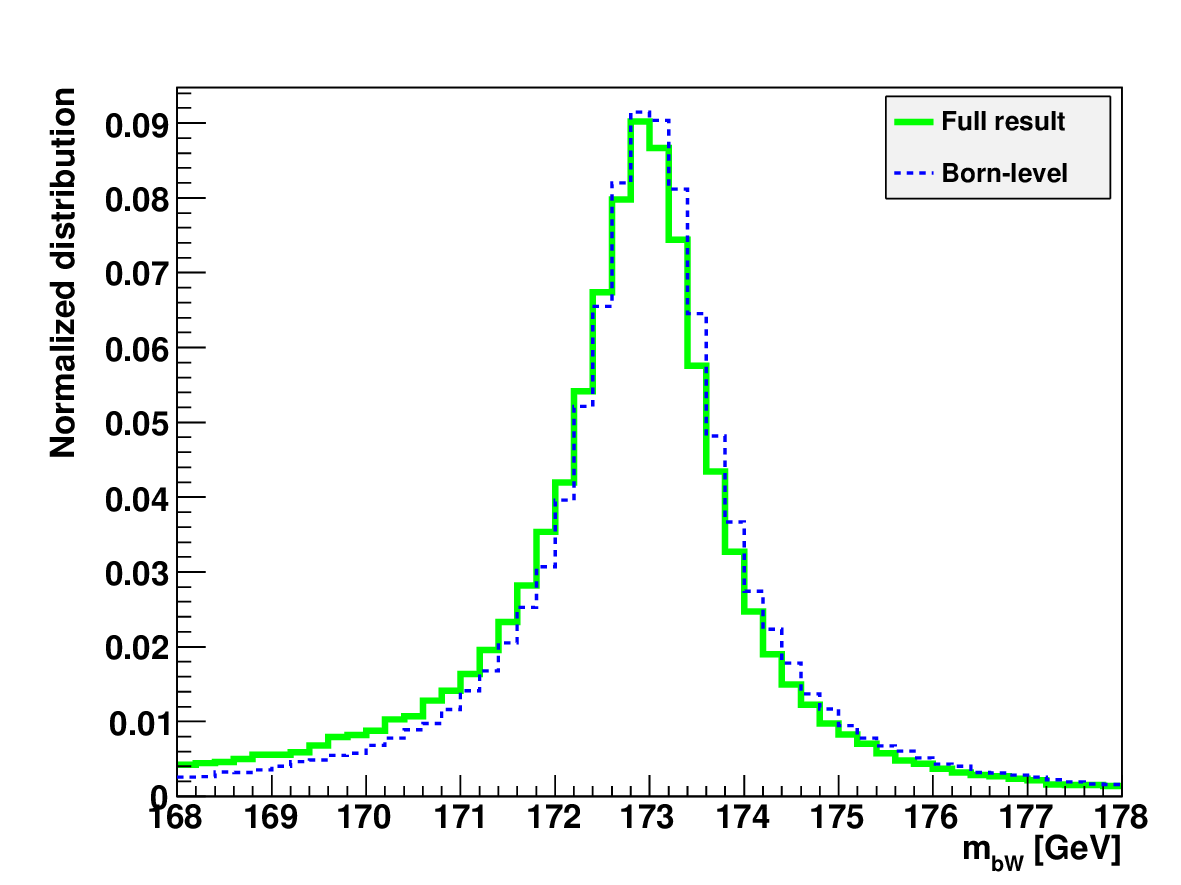}
\\
(a)\hspace{70mm}(b)
\caption{Normalized distributions of (a) the top-quark
 momentum,
 $p=|\vec{p}_t|=|\vec{p}_{b}+\vec{p}_{W^{+}}|$, and 
(b) the $bW$ invariant-mass, 
both defined in the lab.\ frame and 
for the
 events with $\mtt\leq370$~GeV.
 Green solid lines represent our full predictions,
while the blue dashed lines represent the
 Born-level predictions.
\label{fig:hist}}
}

Other top-quark
distributions are less affected by the bound-state effects.
In Fig.~\ref{fig:hist}(a), we show the normalized
distribution of the top-quark momentum
$p=|\vec{p}_t|=|\vec{p}_{b}+\vec{p}_{W^{+}}|$ 
in the laboratory frame.
In Fig.~\ref{fig:hist}(b), we show the normalized
distribution of the invariant-mass of $bW$ 
(${bW^+}$ or $\bb W^{-}$).
The Born-level and full predictions 
are shown by the green solid and blue dashed
lines, respectively.
These lines in Fig.~\ref{fig:hist}(b) correspond to the projections of
Figs.~\ref{fig:mt1mt2}(a,b) to the $m_{bW^+}$ (or $m_{\bb W^-}$)
axis.
All the histograms in Figs.~\ref{fig:hist}(a,b)
are normalized, such that their
integrals take the same value.

In $e^+e^-$ collisions, the top-quark momentum distribution 
in the threshold region is known to
be proportional to the absolute square of the momentum-space Green
function~\cite{Sumino:1992ai,Jezabek:1992np}, whose
shape is strongly influenced by the bound-state effects.
At hadron colliders,
the top-quark momentum is
boosted along the beam direction, and also the partonic collision energy
is not fixed.
As a result, even if we limit the events to those with
$\mtt\leq370$~GeV, the distribution of the top-quark
momentum ($p$, defined in the lab.\ frame) 
is not much affected 
by the bound-state effects at hadron
colliders.

The $m_{bW}$ distribution is important for the determination of the
top-quark mass, hence it should be understood well.
The bound-state effects  deform the Born-level
$m_{bW}$ distribution 
towards the lower side.
The mean values of $m_{bW}$
over the range $|m_{bW}-\mt|<5$~GeV 
are estimated to be 172.7~GeV and 172.9~GeV, 
for the full and Born-level predictions, respectively.
The change of the mean value is about $-200$~MeV,
for the restricted events with $\mtt\leq370$~GeV.
At the LHC 7~TeV, we obtain almost the same result as in 
the 14~TeV case.
At the Tevatron 1.96~TeV, where the $\qqb$ color-octet channel
dominates, the mean values of $m_{bW}$ are estimated as
172.96~GeV and
172.98~GeV, respectively.
Thus, the variation of the mean value is rather small.
Note that {\tt MC@NLO} predicts a $m_{bW}$ distribution similar to the
Born-level distribution, since it simply re-weights the on-shell $\ttb$
cross-section by skewed Breit-Wigner functions.

\section{Summary}

In the first part of this paper
(Sec.~2), we explain our theoretical framework
for including the bound-state effects in the
fully differential cross sections for
the top-quark production and their
subsequent decay processes at hadron colliders.

We formulate a theoretical basis to compute
the fully differential top-quark
cross-sections, which are valid at leading-order both
in the threshold and high-energy regions, and
which smoothly interpolate between the two regions.
The tree-level $\ttb$ double-resonant amplitude
for each process is 
multiplied by a  correction factor, which is written
in terms of the (well-known) non-relativistic Green function, 
but using a modified energy.
This prescription preserves the required unitarity
relation between the total and differential cross sections, 
which would be seriously violated had we
used the naive non-relativistic formula for the
differential cross sections at higher energies.

We also include into the cross sections
two important subleading corrections
induced by the large top-quark width.
(i) In addition to the $\ttb$ double-resonant diagrams,
which receives the above bound-state corrections, we include
the contributions of non-resonant diagrams, whose effects  
are comparatively larger at higher energies.
(This is more or less trivial.)
(ii) As long as we perform numerical integrations of
the differential cross sections,
a sizable phase-space-suppression effect
in the threshold region is inevitable, due to 
the sizable off-shellness of 
top quarks.
In order to effectively account for the gauge cancellation 
by the Coulomb enhancement,
we 
compensate the phase-space-suppression effect (by hand).

Finally we incorporate ISR and the $K$-factors.
ISR effects are incorporated by connecting our differential 
cross sections to a parton-shower simulator such as
{\tt PYTHIA} or {\tt HERWIG}.
We determine the $K$-factors for the cross sections
in the individual channels by matching them to the 
corresponding NLO $\ttb$ invariant-mass distributions
in the threshold region.
With an appropriate choice of scales and 
using $\mtt$-independent $K$-factors, 
we have checked that the color-summed
$\ttb$ invariant-mass distribution agrees with the 
conventional NLO prediction by {\tt MC@NLO} 
reasonably well at high energies.
\medbreak

In the latter part of the paper (Sec.~3),
using the above fully differential cross sections, 
we compute numerically various kinematical distributions of the top
quark, constructed from the momenta of the
$\bwbw$ final state (at the parton level).
Our computations are carried out by MC
event-generation using
{\tt MadGraph}, after implementing the color decomposition
and the bound-state corrections to the output codes.

We confirmed that our prediction reproduces the known NLO predictions
for the $\ttb$ invariant-mass distribution in the threshold region
(which include bound-state effects) at the LHC 7~TeV or 14~TeV; in
particular it exhibits the $1S$ resonance peak below threshold.
Furthermore, our prediction
approaches smoothly to the conventional NLO prediction
(without bound-state effects)
at higher invariant-masses,
from around 30~GeV above the threshold.

We restrict the events to those with
$\mtt\leq 370$~GeV (in the case $2\mt=346$~GeV),
corresponding to about 10\% of the full events, 
and examine 
kinematical distributions other than the
$\mtt$ distribution.
In particular, a characteristic bound-state effect on the
$(bW^+)$-$(\bar{b}W^-)$ double-invariant-mass distribution
is observed.
The distribution is deformed from the 
double Breit-Wigner shape, towards the configuration in
which one of the $\ttb$ pair  is close to on-shell and the other 
has a smaller invariant-mass than $\mt$.

The effect can be understood as a consequence of a
competition between the contributions from
the (dominant) color-singlet
Green function and from the $t$ and $\bar{t}$
propagators.
If the top-quark width were tiny, the Breit-Wigner distribution would
tend to a delta-function, and the top quarks would be forced to
on-shell.
Due to the large decay width, however, 
the binding effect (towards off-shell mass)
and the Breit-Wigner constraint (towards on-shell mass)
remain to be competitive up to a few tens GeV above the threshold.
This effect lowers the mean value of each $bW$ invariant-mass
by a few hundred MeV for the above restricted set of
events.
The correlated deformation of the double-invariant-mass
distribution may affect the mass cut and eventually the top
quark mass measurement at the LHC.
This requires further
careful investigations.
It would be worth emphasizing that the bound-state
effect elucidated here
can never be seen in the conventional perturbative
QCD corrections to the {\it on-shell} $\ttb$ productions,
since the off-shellness of the top quark by the LO 
Coulomb binding effects
plays a crucial role,
and therefore it signifies a unique aspect of the present study.

We examine other distributions, namely the
(single) $bW$ invariant-mass distribution and top-quark-momentum
distribution.
The bound-state effects on these distributions
as a whole are not very significant, although there are certain
systematic tendencies
in the small deformations of the distributions, such as
the aforementioned shift
of the mean value of the invariant-mass.
Furthermore, the dilepton distributions are examined 
 including the leptonic decays of $W$'s at
the matrix-element level (Appendix~D).
We have confirmed the previous observations that the effects of
$W$-boson polarization are quite significant, whereas we find
that the bound-state effects are much smaller.

\acknowledgments
We thank K.~Hagiwara for a fruitful collaboration at an early stage of
this work.
We are also grateful to R.~Frederix, J.~Kanzaki and F.~Maltoni for useful
discussions and comments.
The work of Y.S.\ is supported in part by Grant-in-Aid for scientific
research No.\ 20540246 from MEXT, Japan.

\clfn

\appendix
\section{Green function of the $\ttb$ system}\label{app:green}

In this Appendix we explain how part of the Feynman amplitude
corresponding to $I\to\ttb\to\bwbw$ ($I=gg$ and $\qqb$) can
be identified with a Green function that dictates the time evolution of
the $\ttb$ system.
The argument is not restricted to the non-relativistic region.
In order to define the bound-state as an eigenstate of the full
Hamiltonian, we consider the limit $\Gt\to 0$.\footnote{%
We are not aware how to incorporate the resonance width in the
formulation explained in this Appendix.
It should not be a problem, however, since we want to find an expression
that is valid as $\Gt\to 0$.
}
\medbreak

Before providing a general argument, 
it would be pedagogical to demonstrate a decomposition
of the three-point function
$\bra{0}T \!:\!\bar{\psi}(x)\gamma^\mu \psi(x)\!:\! \,\bar{\psi}(y)\,
\psi(z) \ket{0}$
in the free top-quark case, i.e., in the limit $\alpha_s\to 0$.
Here, $\psi(x)$ represents the top-quark field operator.
The three-point function is included, for example, in the 
amplitude for $\gamma^*\to\ttb\to\bwbw$.
One can express the three-point function as
\bea
&&
\bra{0}T \!:\!\bar{\psi}(x)\gamma^\mu \psi(x)\!:\! \,\bar{\psi}(y)\,
\psi(z) \ket{0}_{\rm tree}
\stackrel{\rm F.T.}{=\!=}
\frac{{p}_t^\alpha\gamma_\alpha+\mt}{{p}_t^2-\mt^2+i\epsilon}
\,\gamma^\mu\,
\frac{-p_{\tb}^\beta\gamma_\beta+\mt}{p_{\tb}^2-\mt^2+i\epsilon}
\rule[-7mm]{0mm}{6mm} \label{free3ptfn1}
\\ &&
~~~~~
=
\Biggl[\frac{\Lambda_+(\vec{p})}{p_{t}^0-\omega(\vec{p})+ i\epsilon}
- \frac{\Lambda_-(\vec{p})}{{p}_{t}^0+\omega(\vec{p})- i\epsilon}\Biggr]
\,\gamma^\mu\,
\Biggl[\frac{\Lambda_-(\vec{p})}{p_{\tb}^0-\omega(\vec{p})+ i\epsilon}
- \frac{\Lambda_+(\vec{p})}{p_{\tb}^0+\omega(\vec{p})- i\epsilon}\Biggr]
\,,
\nn
\eea
where F.T.\ stands for
the Fourier transform, and
$\omega(\vec{p})=\sqrt{\vec{p}^{\,2}+\mt^2}$.
In the second equality,
we have separated the contributions of particle propagating forward
in time and antiparticle propagating backward in time.
Here,
\bea
\Lambda_\pm(\vec{p})=
\frac{\pm\omega(\vec{p})\gamma^0-\vec{p}\cdot\vec{\gamma}+ \mt}
{2\omega(\vec{p})}
\eea
represent the projection operators of a four-component spinor to the
$t$ and $\tb$ components.
Let us drop the contributions of non-resonant parts (far off-shell
contributions) in the last line of Eq.~(\ref{free3ptfn1}), which are
${\cal O}(\Gt/\mt)$ had we retained the top-quark width in the top
quark propagator.
This corresponds to taking the contribution of the time ordering
$x^0<y^0,z^0$ of the left-hand-side of the equation.
Noting that $p_t^0+p_{\tb}^0=\mtt$, the resonant part can
be expressed as
\bea
&&
\frac{\Lambda_-(\vec{p})}{p_{\tb}^0-\omega(\vec{p})+ i\epsilon}
\,\gamma^\mu\,
\frac{\Lambda_+(\vec{p})}{p_{t}^0-\omega(\vec{p})+ i\epsilon} 
\rule[-7mm]{0mm}{6mm}
\nn\\ &&
~~~~~
=\frac{\Lambda_-(\vec{p})\,\gamma^\mu\,\Lambda_+(\vec{p})}
{\mtt-2\omega(\vec{p}) + i\epsilon}
\times
\Biggl[ \frac{1}{p_{t}^0-\omega(\vec{p})+ i\epsilon} +
\frac{1}{p_{\tb}^0-\omega(\vec{p})+ i\epsilon} \Biggr] .
\label{free3ptfn}
\eea
In coordinate space, this equation corresponds to
the splitting of the time ordering
$x^0<y^0,z^0$ into two orderings 
$x^0<y^0<z^0$  and $x^0<z^0<y^0$.
On the right-hand side,
the factor outside the square bracket represents
the time evolution of the $\ttb$ system.
We may identify the denominator of this factor with
$\mtt-H_0+i\epsilon$ and
observe that the Hamiltonian $H_0$ of the free $\ttb$ system
is given by $2\omega(\vec{p})=2\sqrt{\vec{p}^{\,2}+\mt^2}$.
The first term in the square bracket
represents the propagation
of $t$ after $\tb$ decayed first, whereas the second term
represents the propagation of $\tb$
after ${t}$ decayed first. 
\medbreak

Now we develop a general argument.
The four-point function of $t$ and $\tb$
\bea
\bra{0}T\, \psi(x_1) \bar{\psi}(x_2) \bar{\psi}(x_3) \psi(x_4) \ket{0}
\eea
is a building block of the Feynman amplitude for $I\to\ttb\to\bwbw$.
The above four-point function can be decomposed to the sum of the
different time orderings of $x_1,x_2,x_3,x_4$.
As shown in \cite{LMT}, the bound-state contributions are included in the
orderings in which ${\rm Min}(x_1^0,x_2^0)>{\rm Max}(x_3^0,x_4^0)$.
In fact, one may use the relation
\bea
\theta(x^0)\stackrel{\rm F.T.}{=\!=}\frac{i}{k^0+i\epsilon} ,
\eea
to show that
\bea
&&
\bra{0}T\, \psi(x_1) \bar{\psi}(x_2) \bar{\psi}(x_3) \psi(x_4) \ket{0}
\biggr|_{{\rm Min}(x_1^0,x_2^0)>{\rm Max}(x_3^0,x_4^0)}
\label{analytic-4ptfn}
\\&&~~~
~~~~~
\stackrel{\rm F.T.}{=\!=} 
\Biggl[ \sum_n \frac{1}{2M_n}\,
\frac{\varphi_n(\vec{p}_f)\bar{\varphi}_n(\vec{p}_i)}
{\mtt-M_n + i\epsilon}
\Biggr]
\nn\\&&~~~~~~~~~~~
\times 
\Biggl[ \frac{1}{p_{t,f}^0-\omega(\vec{p}_f)+ i\epsilon} + 
\frac{1}{p_{\tb,f}^0-\omega(\vec{p}_f)+ i\epsilon} \Biggr]
\nn\\&&~~~~~~~~~~~
\times 
\Biggl[ \frac{1}{p_{t,i}^0-\omega(\vec{p}_i)+ i\epsilon} + 
\frac{1}{p_{\tb,i}^0-\omega(\vec{p}_i)+ i\epsilon} \Biggr]
+(\mbox{non-resonant part})
,
\nn\eea
in the c.m.\ frame of the $\ttb$ pair.
$p_{t,i}^\mu$ and $p_{t,f}^\mu$ ($p_{\tb,i}^\mu$ and
$p_{\tb,f}^\mu$) denote the four momenta of the initial and final
$t$ ($\tb$), respectively, in the $\ttb$ c.m.\ frame;
$\vec{p}_i=\vec{p}_{t,i}=-\vec{p}_{\tb,i}$ and
$\vec{p}_f=\vec{p}_{t,f}=-\vec{p}_{\tb,f}$.
The first line on the right-hand side of Eq.~(\ref{analytic-4ptfn})
is identified with the Green function of the $\ttb$ system,
which includes the bound-state poles,
$\bra{\vec{p}_f}[\mtt-H+i\epsilon]^{-1}\ket{\vec{p}_i}$.
The bound-state wave functions are defined as\footnote{%
$\varphi_n(\vec{p})$ is related to the Bethe-Salpeter wave function 
$\chi_n(p)$ by 
\bea
&&
\chi_n\biggl(\frac{p_t-p_{\tb}}{2}\biggr)
\equiv
\int d^4(x-y)\, \exp\biggl[i \biggl\{
\frac{p_t-p_{\tb}}{2} \cdot (x-y) +  M_n \frac{x^0+y^0}{2}
\biggr\}\biggr]\,
\bra{0}T\, \psi(x)\,\bar{\psi}(y) \ket{n}
\nn\\&&
~~~~~
=
\varphi_n(\vec{p})
\biggl[\frac{i}{p_t^0-\omega(\vec{p})+i\epsilon}+
\frac{i}{p_{\tb}^0-\omega(\vec{p})+i\epsilon}\biggr] 
+(\mbox{terms without a single particle pole}) .
\eea
}
\bea
&&
\varphi_n(\vec{p})=
\frac{\bra{0}\psi(0)\ket{t;\vec{p}}\bra{t;\vec{p}}\bar{\psi}(0)
\ket{n}}{2\omega(\vec{p})}= \pm
\frac{\bra{0}\bar{\psi}(0)\ket{\tb;-\vec{p}}
\bra{\tb;-\vec{p}}{\psi}(0)
\ket{n}}{2\omega(\vec{p})},
\nn\\&&
\bar{\varphi}_n(\vec{p})=
\frac{\bra{n}\psi(0)\ket{{t};\vec{p}}\bra{{t};\vec{p}}\bar{\psi}(0)
\ket{0}}{2\omega(\vec{p})}= \pm
\frac{\bra{n}\bar{\psi}(0)\ket{\tb;-\vec{p}}
\bra{\tb;-\vec{p}}{\psi}(0)
\ket{0}}{2\omega(\vec{p})} .
\eea
$|\!\!\stackrel{(-)}{t};\vec{p}\rangle$ denotes the (anti)top-quark
one-particle state with momentum $\vec{p}$.
The bound-state $\ket{n}$ is defined as an eigenstate
of the full Hamiltonian, $H\ket{n}=M_n\ket{n}$, and it is
assumed to be a $CP$ eigenstate.

In Eq.~(\ref{analytic-4ptfn}), the bound-state poles stem from the time
evolution between ${\rm Min}(x_1^0,x_2^0)$ and ${\rm Max}(x_3^0,x_4^0)$,
whereas the single particle poles stem from the time evolution between
$x_1^0$ and $x_2^0$ and between $x_3^0$ and $x_4^0$.
For instance, in the case that $x_1^0>x_2^0>x_3^0>x_4^0$, one may insert
the projection operators to the subspaces spanned by single particle
states,
$\int\frac{d^3\vec{p}}{(2\pi)^32\omega(\vec{p})}
\ket{t;\vec{p}}\bra{t;\vec{p}}$
and $\int \frac{d^3\vec{p}}{(2\pi)^3 2\omega(\vec{p})}
\ket{\tb;-\vec{p}} \bra{\tb;-\vec{p}}$,
to extract the contributions from single-particle poles.
Then the bound-state poles and single-particle poles appear from the
Fourier transform
\bea
&&
\theta(x_2^0-x_3^0) \,
\bra{t;\vec{p}} \bar{\psi}(0,\vec{x}_2)\,
e^{-iH(x_2^0-x_3^0)} \,
 \bar{\psi}(0,\vec{x}_3)  \ket{\tb;-\vec{p}}
\rule[-3mm]{0mm}{6mm}
\label{theeq}
\\&&
\stackrel{\rm F.T.}{=\!=} 
\bra{t;\vec{p}} \bar{\psi}(0)\,
\frac{1}{\mtt-H + i\epsilon} \,
 \bar{\psi}(0)  \ket{\tb;-\vec{p}}
\rule[-3mm]{0mm}{6mm}
 \nn\\&&
 =
 \sum_n \frac{1}{2M_n}\,
\frac{\bra{t;\vec{p}} \bar{\psi}(0)\ket{n}
\bra{n}\bar{\psi}(0)\ket{\tb;-\vec{p}}}
{\mtt-M_n + i\epsilon}
+ (\mbox{non-resonant part}) ,
 \nn\eea
and
\bea
&&
\theta(x_1^0-x_2^0)\,
\bra{0}\psi(0)\,e^{-iH(x_1^0-x_2^0)}\ket{t;\vec{p}}
\\&&~~~~~~~~~~~~~~~~~~~
\stackrel{\rm F.T.}{=\!=} 
\frac{i}{p_t^0-\omega(\vec{p})+i\epsilon}
\bra{0}\psi(0)\ket{t;\vec{p}}
\nn\\ &&
~~~~~~~~~~~~~~~~~~~~~~~~
+(\mbox{terms without a single particle pole}) .
 \nn\eea
We may also express $\theta(x_3^0-x_4^0)\,
\bra{\tb;-\vec{p}}e^{-iH(x_3^0-x_4^0)}\,\psi(0)\ket{t;\vec{p}}
$
in a similar manner.
The second line of Eq.~(\ref{theeq}) can be identified with the
Green function of the $\ttb$ system.

\section{Derivation of the off-shell suppression
 effect}\label{app:off}

In this appendix we derive the off-shell suppression effect
for the process $I\to \ttb$ with $I=gg$ or $I=\qqb$ in the
threshold region.
A similar formula for the process $e^+e^-\to\ttb$ was derived
in~\cite{Sumino:1992ai}.

The tree-level double-resonant amplitude has a form
\bea
&&
 \cM_{\ttb}^{(c)}(p_1,p_2;p_{b},p_{W^+},p_{\bb},p_{W^-})_{\rm tree} \\
&&~~~~~
 =  \cD(p_{t};p_{b},p_{W^+})\cdot \cS(p_{t})\cdot
 \cP^{(c)}_I(p_1,p_2;p_{t},p_{\tb})\cdot
 \cSb(p_{\tb})\cdot \bar\cD(p_{\tb};p_{\bb},p_{W^-}),
 \nn\eea
where $\cD$ and $\bar{\cD}$ represent the decay part of $t$ and $\tb$,
respectively,
$\cS$ and $\cSb$ denote the propagators for $t$ and $\tb$, respectively,
in FWS:
\begin{subequations}
\bea
&& 
\cS(p_{t})=
 \frac{i(p\dsl{-5pt}_{t}+\mt)}{p_{t}^{2}-\mt^{2}+i\mt\Gt},\\
 &&
 \cSb(p_{\tb})=
 \frac{i(-p\dsl{-5pt}_{\tb}+\mt)}{p_{\tb}^{2}-\mt^{2}+i\mt\Gt}\,.
\label{eq:sf}
\eea
\end{subequations}
$\cP^{(c)}_{I}$ represents the $\ttb$ production part
which depends on the initial-state partons $I$.

Integrating $|\cM^{(c)}_{\ttb,{\rm tree}}|^2$
over the $\bwbw$ phase-space for a fixed $\ttb$ invariant-mass 
$\sqrt{s}$, we
obtain the $\ttb$ invariant-mass distribution in the following form:
\begin{align}
 \hat\sigma^{(c)}_{I\to\bwbw}(s) =
 \int\frac{ds_t}{2\pi}\frac{ds_{\tb}}{2\pi}
 \rho(s_t)\rho(s_{\tb})\,\hat\sigma^{(c)}_{I,{\rm off}}(s;s_t,s_{\tb}),
 \label{eq:off}
\end{align}
where 
\bea
\rho(s)=2\sqrt{s}\Gt(s)|\Delta_{F}(s)|^2\,,
~~~~~~~~
\Delta_{F}(s)=\frac{1}{s-\mt^2+i\mt\Gt}\,,
\eea
and
$\Gt(s)$
denotes the running width of the top quark with
$p_t^2=s$ in the unitary
gauge.
Explicitly it is given by
replacing $\mt^2$ by $s$ in the on-shell decay width: 
\begin{align}
 \Gt(s) = \theta\left(s-(\mW+\mb)^2\right)
 \frac{G_{F}s^{\frac{3}{2}}}{8\sqrt{2}\pi}
 \lambda(1,\frac{\mW^2}{s},\frac{\mb^2}{s})
 f(\frac{\mW^2}{s},\frac{\mb^2}{s})\,,
 \label{Gt-unitarygauge}
\end{align}
where $f(x,y)=1+x-2y-2x^2+xy+y^2$ and
$\lambda(a,b,c)=\sqrt{a^2+b^2+c^2-2(ab+bc+ca)}$.
In Eq.~(\ref{eq:off}),
$\hat\sigma_{i,{\rm off}}^{(c)}
(s;s_t,s_{\tb})$ represents the off-shell $\ttb$ cross-section,
corresponding to the $t$ and $\tb$
invariant-masses of $s_{t}$ and $s_{\tb}$, respectively.

The formula (\ref{eq:off}) is obtained in the following manner.
First we define the $t$ and $\tb$
decay parts of the matrix-element squared, integrated over
the $bW$ phase-space,  as
\begin{subequations}
\bea
&&
 \sigma_{t}(p\dsl{-6pt}_{t}) \equiv \int d\Phi_{2}(p_{t}^2;p_b,p_{W^+})
 \sum \left(\cD \cD^{\mathfrak{d}}\right) =
 \frac{\Gt(p_t^2)}{\sqrt{p_t^2}}p\dsl{-6pt}_{t}(1-\gamma_{5}),
\\&&
 \bar\sigma_{\tb}(p\dsl{-6pt}_{\tb}) \equiv \int
 d\Phi_{2}(p_{\tb}^2;p_{\bb},p_{W^-})\sum \left(\bar{\cD}^{\mathfrak{d}}
 \bar{\cD}\right) =
 \frac{\Gt(p_{\tb}^2)}{\sqrt{p_{\tb}^2}}p\dsl{-6pt}_{\tb}(1-\gamma_{5}),
\eea\label{eq:dfunc}
\end{subequations}
where the summation is over the spins of the final $b$-quark and
$W$-boson.
Here and hereafter, $A^\mathfrak{d}=\gamma^0A^{\dagger}\gamma^0$
denotes the Dirac conjugate.
Decomposing the four-body
phase-space and utilizing Eqs.~(\ref{eq:dfunc}), the $\bwbw$
cross-section is given by
\begin{align}
 \hat\sigma^{(c)}_{I\to\bwbw}(s) =
 \frac{1}{2s}\int\frac{dp_{t}^2}{2\pi}\frac{dp_{\tb}^2}{2\pi}
 d\Phi_2(s;p_t,p_{\tb})\,{\rm \overline{Tr}}\left[
 \left\{\cS^{\mathfrak{d}}\sigma_{t}\cS\right\}\cdot\cP_{I}^{(c)}\cdot
 \left\{\cSb\bar\sigma_{\tb}\cSb^{\mathfrak{d}}\right\}\cdot
 \cP^{(c){\mathfrak{d}}}_{I}\right],
\end{align}
where ${\rm \overline{Tr}}$ includes averaging 
over the spins and colors of
the initial-state partons.
The spinor matrices in the curly brackets
$\{\cdots\}$ are calculated to be
\begin{subequations}
\begin{align}
 \cS^{\mathfrak{d}}\sigma_{t}\cS&=
 2\sqrt{p_t^2}\,\Gt(p_t^2)|\Delta_{F}(p_{t}^2)|^2
 \Sigma_{t}(p\dsl{-6pt}_t),\\
 \cSb\bar\sigma_{\tb}\cSb^{\mathfrak{d}} &=
 2\sqrt{p_{\tb}^2}\,\Gt(p_{\tb}^2)|\Delta_{F}(p_{\tb}^2)|^2
 \Sigma_{\tb}(p\dsl{-6pt}_{\tb}),
\end{align}
\end{subequations}
with 
\begin{align}
 \Sigma_{t,\tb}(p\dsl{-6pt}) =
 \frac{p^2+\mt^2}{2p^2}p\dsl{-6pt} \pm \mt
 + \frac{p^2-\mt^2}{2p^2}p\dsl{-6pt}\gamma_{5}. 
\end{align}
Thus, we obtain Eq.~(\ref{eq:off}) with the off-shell cross
section
\begin{align}
 \hat\sigma^{(c)}_{I,{\rm off}}(s;p_t^2,p_{\tb}^2) = \frac{1}{2s}\int
 d\Phi_2(s;p_t,p_{\tb})\,{\rm \overline{Tr}}\left[
 \Sigma_{t}(p\dsl{-6pt}_t)\cdot\cP_{I}^{(c)}\cdot
 \Sigma_{\tb}(p\dsl{-6pt}_{\tb}) \cdot \cP^{(c){\mathfrak{d}}}_{I} \right].
\end{align}
Note that in the case $p_t^2=p_{\tb}^2=\mt^2$,
$\sigma^{(c)}_{I,{\rm off}}$ 
equals the on-shell $\ttb$ production cross section.

Below the $\ttb$ threshold, 
the dominant kinematical configuration
is such that either one of $t$ and $\tb$ is on-shell,
because of the presence of the $|\Delta_F|^2$ factors.
Taking this into account, the 
off-shell cross section can be approximated by
\begin{align}
 \hat\sigma^{(c)}_{I,{\rm off}}(s;s_t,s_{\tb}) \simeq
{\hat\sigma^{(c)}_{I\to\ttb}(s)}\,
 \frac{\lambda(1,\frac{s_t}{s},\frac{s_{\tb}}{s})}{\bt}\,
 F_{I}\Bigl(\frac{\mt^2}{s_t},\frac{\mt^2}{s_{\tb}}\Bigr),
\end{align}
where $\hat\sigma^{(c)}_{I\to\ttb}(s)$ are the cross sections for the
on-shell $\ttb$ productions; $F_{gg}(x,y)=xy$ and
$F_{\qqb}(x,y)=(2+x+y)/4$.
The factor $\lambda(1,\frac{s_t}{s},\frac{s_{\tb}}{s})$ originates from the
$\ttb$ phase-space volume and reduces to
$\bt=\sqrt{1-\frac{4\mt^2}{s}}$ in the on-shell limit $s_t=s_{\tb}=\mt^2$.

Thus, the ratio of the $\bwbw$ cross section to the on-shell
$\ttb$ cross section is
given by
\begin{align}
 \frac{\sigma^{(c)}_{I\to\bwbw}(s)}{\sigma^{(c)}_{I\to\ttb}(s)} \simeq
 \int \frac{ds_t}{2\pi} \frac{ds_{\tb}}{2\pi}\,
 \rho(s_t)\rho(s_{\tb})\,
 \frac{\lambda(1,\frac{s_t}{s},\frac{s_{\tb}}{s})}{\bt}\,
 F_{I}\Bigl(\frac{\mt^2}{s_t},\frac{\mt^2}{s_{\tb}}\Bigr).\label{eq:bwbwtt}
\end{align}
$\sqrt{s_t}\Gt(s_t)\sqrt{s_{\tb}}\Gt(s_{\tb})$
in the $\rho$ factors, which stems from the $bW$ phase-space volumes,
possesses strong dependences on the invariant-masses of $t$ and $\tb$:
$\sqrt{s}\Gt(s)\sim{s}^{2}$ since the running width behaves as
$\Gt(s)\sim{s}^{{3}/{2}}$.
Below the threshold, either of $t$ and $\tb$ is forced to be off-shell
($p^2<\mt^2$), hence $\sqrt{s_t}\Gt(s_t)\sqrt{s_{\tb}}\Gt(s_{\tb})$
acts as a suppression factor.
On the other hand, the remaining factors $\lambda/\beta_t$ and $F_I$
give only weak enhancement near the
threshold.
Thus, the effect of phase-space-suppression is approximately
accounted for by the ratio
\bea
\frac{\sqrt{s_t}\Gt(s_t)}{\mt\Gt}\cdot
 \frac{\sqrt{s_{\tb}}\Gt(s_{\tb})}{\mt\Gt}
\eea
for fixed $s_t$ and $s_{\tb}$.
The ratio Eq.~(\ref{eq:bwbwtt}) is considerably less
than unity below the threshold.

\section{Color decomposition of the amplitude and color flow in $gg$
 channel}\label{app:color}

In this section, we describe the color structure of the $gg\to\ttb$
matrix-element, and its decomposition into color-singlet and octet
components.
Moreover, we comment on the color flow in $gg$ channel.

By explicitly stating the color indices of initial-state gluons ($a$,
$b$) and final-state $t$ and $\tb$ ($i$,$j$), the matrix element at the
Born-level can be written in a following form;
\begin{align}
 M^{ab}_{ij}(p_k,\lambda_k) =
 \frac{1}{2}\left\{T^{a},T^{b}\right\}_{ij}M_{S}(p_k,\lambda_k)
 + \frac{1}{2}\left[T^{a},T^{b}\right]_{ij}M_{A}(p_k,\lambda_k),
\label{eq:b1}
\end{align}
where $M_{S}$ and $M_{A}$ represent the subamplitudes for
color-symmetric and anti-symmetric part, respectively, which depend on
the 4-momenta and helicities of initial- and final-state particles.
The color-factor in the color-symmetric part is decomposed into
\begin{align}
 \frac{1}{2}\left\{T^{a},T^{b}\right\}_{ij} =
 \frac{1}{2N}\delta^{ab}\delta_{ij} + \frac{1}{2}d^{abc}T^{c}_{ij},
\end{align}
where the first term represents color-singlet contribution and the
second term color-octet.
The color-anti-symmetric part contains only color-octet contribution;
\begin{align}
 \frac{1}{2}\left[T^{a},T^{b}\right]_{ij}
 = \frac{i}{2}f^{abc}T^{c}_{ij}.
\end{align}
Since each part do not interfere with each other, the absolute square of
the amplitude with summing over colors is given as a sum of each
contribution;
\begin{align}
 \sum_{\rm colors}\left|M^{ab}_{ij}\right|^2 & =
 \tsum\left|\frac{1}{2N}\delta^{ab}\delta_{ij}\right|^2
 \left|M_{S}\right|^2 \nn \\
 & + \tsum\left|\frac{1}{2}d^{abc}T^{c}_{ij}\right|^2
 \left|M_{S}\right|^2
 + \tsum\left|\frac{i}{2}f^{abc}T^{c}_{ij}\right|^2
 \left|M_{A}\right|^2. \label{eq:me2}
\end{align}
The first line in the r.h.s.\ in Eq.~(\ref{eq:me2}) is the color-singlet
contribution and the second line the color-octet. \\

Alternatively, we may express the amplitude in the following basis;
\begin{align}
 M^{ab}_{ij}(p_k,\lambda_k) =
 \left(T^{a}T^{b}\right)_{ij}M_{J_1}(p_k,\lambda_k) +
 \left(T^{b}T^{a}\right)_{ij}M_{J_2}(p_k,\lambda_k).
 \label{eq:b2}
\end{align}
The two basis, Eq.~(\ref{eq:b1}) and Eq.~(\ref{eq:b2}) are related by
$M_{S}=M_{J_1}+M_{J_2}$ and $M_{A}=M_{J_1}-M_{J_2}$.
The absolute square of the amplitude is then written in a matrix form
as,\footnote{%
This matrix corresponds to the matrix {\tt CF} in the {\tt
MadGraph} code ({\tt matrix.f}).}
\begin{align}
 \sum_{\rm colors}\left|M^{ab}_{ij}\right|^2 = \frac{1}{3}
 \left(
 \begin{array}{cc}
  M^{*}_{J_1}&M^{*}_{J_2}
 \end{array}
 \right) \left(
 \begin{array}{rr}
  16& -2 \\ -2 & 16
 \end{array}
 \right) \left(
 \begin{array}{c}
  M_{J_1} \\ M_{J_2}
 \end{array}
 \right).\label{eq:me2m}
\end{align}

By rewriting the absolute square of the color-singlet part of the
amplitude in this basis, the corresponding matrix for the color-singlet
$gg\to\ttb$ amplitude is found to be
\begin{align}
 \tsum\left|\frac{1}{2N}\delta^{ab}\delta_{ij}\right|^2
 \left|M_{S}\right|^2 = \frac{1}{3}
 \left(
 \begin{array}{cc}
  M^{*}_{J_1}&M^{*}_{J_2}
 \end{array}
 \right) \left(
 \begin{array}{rr}
  2 & 2 \\ 2 & 2
 \end{array}
 \right) \left(
 \begin{array}{c}
  M_{J_1} \\ M_{J_2}
 \end{array}
 \right).\label{eq:me2s}
\end{align}
The color-octet contribution is obtained by subtracting
Eq.~(\ref{eq:me2s}) from Eq.~(\ref{eq:me2m}). \\

Finally, we comment on the color flow in $gg$ channel.\footnote{%
We thank F.~Maltoni for pointing out the modification of the color-flow
structure in {\tt MadEvent}, see also Ref.~\cite{Artoisenet:2007qm}.}
In the color-singlet case, the color flow is disconnected between
initial-state and final-state, reflecting the color-factor
$\delta^{ab}\delta_{ij}$; see the left diagram in Fig.~\ref{fig:cf}.

In the color-octet case, there exist two kind of color flows, middle and
right diagrams in Fig.~\ref{fig:cf}, associated with the color-factor; 
$(T^aT^b)_{ij}$ and $(T^bT^a)_{ij}$, respectively.
Either of the two may be selected according to the ratio
$|M_{J_1}|^2:|M_{J_2}|^2$ at each phase-space point and given helicities
of initial and final-state particles in event generations.
The color flow in $\qqb$ channel is unique at Born-level.
\FIGURE[t]{
\includegraphics[width=.8\textwidth,clip]{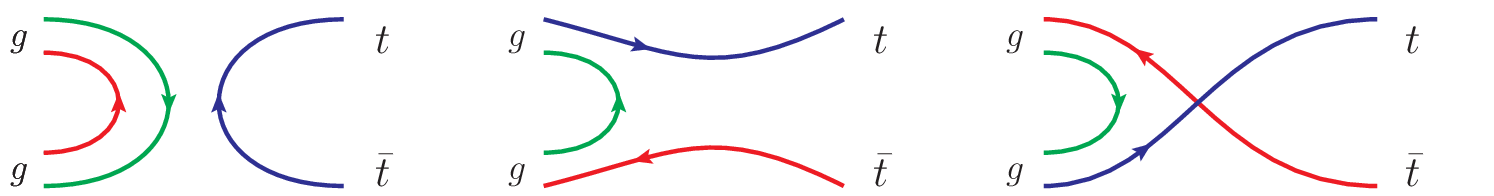}
\caption{
 Color-flow diagrams in the $gg\to\ttb$ amplitudes.
 Left diagram representing the color-factor $\delta^{ab}\delta_{ij}$ is
 for the color-singlet case, middle and right diagrams representing the
 color-factor $(T^aT^b)_{ij}$ and $(T^bT^a)_{ij}$, respectively, are for
 the color-octet case.
 \label{fig:cf}}
}

\section{Dilepton distributions in dilepton decay mode}
\label{app:wdecay}

In the main body of this paper we assume that the $W$-bosons 
from top-quark decays are on-shell.
In this appendix, we take into account decays of the $W$-bosons 
at the matrix-element level.
The advantage
is to correctly take into account the off-shellness and polarization of
the $W$-bosons.
This is crucial to predict correctly the
angular distributions as well
as other kinematical distributions
of the decay daughters of $W$'s.
As an example, we show some distributions in the dilepton decay mode
$pp\to\bwbw\to b\ell^{+}\nu_{\ell}\bb\ell^{-}\bar{\nu}_{\ell}$,
where $\ell=e$, $\mu$.

We calculate the amplitudes from the set of Feynman diagrams obtained by
just adding the $W\to\ell\nu_{\ell}$ vertex to 
the Feynman diagrams for the $\bwbw$ final state.
We generate the dilepton events at the LHC $\sqrt{s}=14$~TeV with
standard kinematical cuts for the lepton momenta, $|\eta_{\ell}|\leq
2.5$ and $p_{T,\ell}\geq 10$~GeV.
To avoid a singularity due to the vanishing running top-quark
width, see Eq.~(\ref{Gt-unitarygauge}) in Appendix~\ref{app:off},
we restrict the phase-space integral region to $m_{bW}>\mb+\mW$.
Note that the matrix-elements for $m_{bW}<\mb+\mW$ are very
suppressed.
We set the $W$-boson decay-width to $\GW=2.05$~GeV.

\FIGURE[t]{
\includegraphics[width=.9\textwidth,clip]{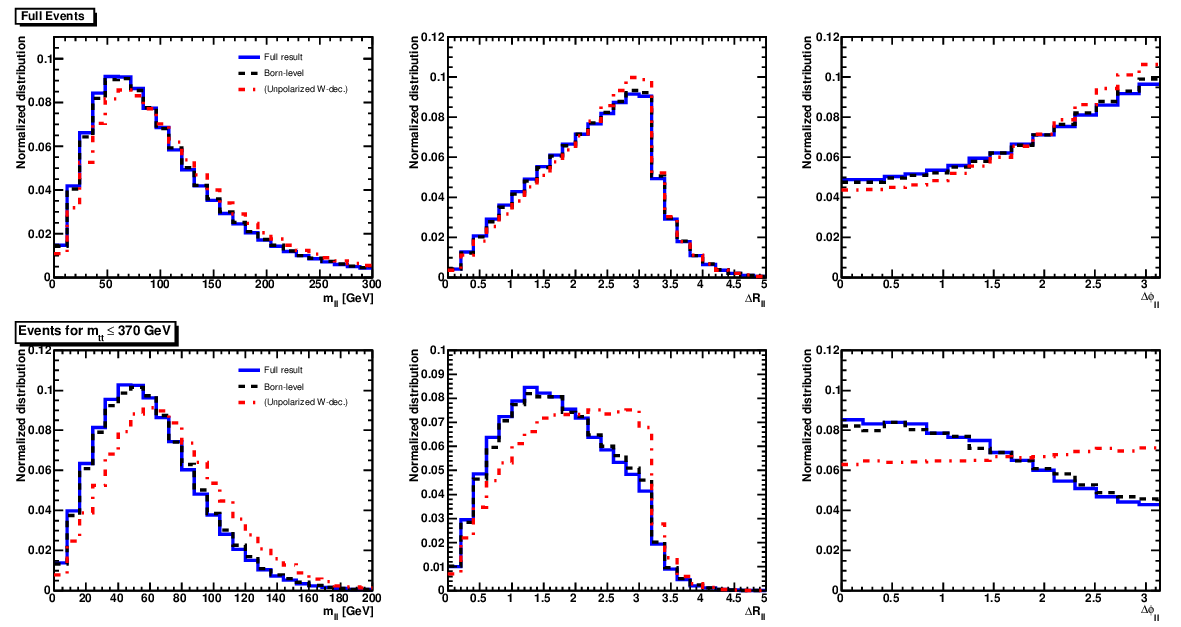}
\caption{Distributions in an invariant-mass (left), a distance
 in $\eta$-$\phi$ plane (middle), and a difference in the azimuthal
 angle (right) of the two leptons for the dilepton events at the LHC
 $\sqrt{s}=14$~TeV.
 Top three graphs are for the full events, and bottom three graphs are
 for the events with $\mtt\leq370$~GeV.
 Blue solid line is the full result, blue dashed line is the Born-level
 result, and red dot-dashed line is obtained by the full calculation but
 assuming unpolarized $W$-bosons.
 \label{fig:dilep}}
}
In Figs.~\ref{fig:dilep}, we plot three different distributions of 
kinematical variables constructed from the four-momenta of
dileptons:
(a1, a2) the invariant-mass of the two leptons, (b1, b2)
the distance in the $\eta$-$\phi$ plane, 
$\Delta{R}_{\dilep}=\sqrt{\Delta\eta_{\dilep}^2+\Delta\phi_{\dilep}^2}$,
and (c1, c2) the difference of the azimuthal angles,
$\Delta\phi_{\dilep}$.
The first three graphs (a1, b1, c1) 
correspond to the events from all the $\mtt$ region,
while
the last three (a2, b2, c2)
to the events with $\mtt\leq370$~GeV.
The solid lines represent our full prediction, 
and the dashed lines represent the Born-level
predictions.
To see the effects of the non-zero $W$-boson polarization,
we also plot the distributions computed from the $\bwbw$
events followed by the leptonic decays of on-shell unpolarized 
$W$-bosons.

We find that, due to the bound-state corrections, all three distributions
are shifted to the lower side, although
the variations are fairly small even for the
events with $\mtt\leq370$~GeV.
By contrast, the effects of
the non-zero $W$-boson polarization is pronounced, 
especially for the events with $\mtt\leq370$~GeV.
The most evident difference can be seen in the $\Delta\phi_{\dilep}$
distribution, where the full calculation predicts that the number of
events at $\Delta\phi_{\dilep}=0$ is more than twice than the number of
events at $\Delta\phi_{\dilep}=\pi$, while almost flat distribution by
assuming the unpolarized $W$-bosons.

This finding is consistent with the similar study in
Ref.~\cite{Mahlon:2010gw} where the importance of the top-quark spin
correlation in the $\Delta\phi_{\dilep}$ distribution is examined for
the events with $\mtt\leq 400$ GeV.
They compare the $\Delta\phi_{\dilep}$ distribution fully taking into
account the top-quark spin correlation with that assuming the spherical
top-quark decay into $bW$'s followed by the correlated $W$-boson decay.
Our calculation assuming unpolarized $W$-boson decays includes the
correct angular distributions in $t\to bW$ decays, but forcing spherical
distributions in $W\to\ell\nu_{\ell}$ decays.
Thus, to predict the dilepton observables, all the spin correlations in
decays of top-quarks and also $W$-bosons are required.
We confirm their finding that the difference in $\Delta{R}_{\dilep}$
distribution comes from mainly the difference in $\Delta\phi_{\dilep}$
distribution.


\end{document}